# Cavity as a source of conformational fluctuation and high-energy state: High-pressure NMR study of a cavity-enlarged mutant of T4 lysozyme

Akihiro Maeno,[†‡] Daniel Sindhikara,[§] Fumio Hirata,[¶] Renee Otten,[&] Frederick W. Dahlquist,[△] Shigeyuki Yokoyama,[††&&] Kazuyuki Akasaka,[†‡] Frans A. A. Mulder,[§§] and Ryo Kitahara[‡¶¶*]

[†]*High Pressure Protein Research Center, Institute of Advanced Technology, Kinki University, 930 Nishimitani, Kinokawa, Wakayama 649-6493, Japan*
[‡]*RIKEN SPring-8 Center, 1-1-1 Kouto, Sayo-cho, Sayo-gun, Hyogo 679-5148, Japan*
[§]*College of Science and Engineering, Ritsumeikan University, 1-1-1 Noji-higashi, Kusatsu, Shiga 525-8577, Japan*
[¶]*College of Life Sciences, Ritsumeikan University, 1-1-1 Noji-higashi, Kusatsu, Shiga 525-8577, Japan*
[&]*Department of Biochemistry, Brandeis University, 415 South Street, Waltham, MA 02454, USA*
[△]*The Department of Chemistry and Biochemistry and the Department of Molecular, Cellular and Developmental Biology, University of California Santa Barbara, Santa Barbara CA 93106-6105, USA*
[††]*RIKEN Systems and Structural Biology Center, 1-7-22, Suehiro-cho, Tsurumi, Yokohama 230-0045, Japan*
[&&]*Department of Biophysics and Biochemistry and Laboratory of Structural Biology, Graduate School of Science, The University of Tokyo, Bunkyo-ku, Tokyo 113-0033, Japan.*
[§§]*Department of Chemistry and Interdisciplinary Nanoscience Center iNANO, University of Aarhus, Gustav Wieds Vej 14, DK-8000 Aarhus C, Denmark*
[¶¶]*College of Pharmaceutical Sciences, Ritsumeikan University, 1-1-1 Noji-higashi, Kusatsu, Shiga 525-8577, Japan*

[*]**Corresponding author:** Ryo Kitahara
E-mail: ryo@ph.ritsumei.ac.jp
Phone: +81-77-561-5751; Fax: +81-77-561-2659

ABSTRACT

Although the structure, function, conformational dynamics, and controlled thermodynamics of proteins are manifested by their corresponding amino acid sequences, the natural rules for molecular design and their corresponding interplay remain obscure. In this study, we focused on the role of internal cavities of proteins in conformational dynamics. We investigated the pressure-induced responses from the cavity-enlarged L99A mutant of T4 lysozyme, using high-pressure NMR spectroscopy. The signal intensities of the methyl groups in the $^{1}$H/$^{13}$C HSQC spectra, particularly those around the enlarged cavity, decreased with the increasing pressure, and disappeared at 200 MPa, without the appearance of new resonances, thus indicating the presence of heterogeneous conformations around the cavity within the ground state ensemble. Above 200 MPa, the signal intensities of more than 20 methyl groups gradually decreased with the increasing pressure, without the appearance of new resonances. Interestingly, these residues closely matched those sensing a large conformational change between the ground- and high-energy states, at atmospheric pressure. $^{13}$C and $^{1}$H NMR line-shape simulations showed that the pressure-induced loss in the peak intensity could be explained by the increase in the high-energy state population. In this high-energy state, the aromatic side chain of F114 gets flipped into the enlarged cavity. The accommodation of the phenylalanine ring into the efficiently packed cavity may decrease the partial molar volume of the high-energy state, relative to the ground state. We suggest that the enlarged cavity is involved in the conformational transition to high-energy states and in the volume fluctuation of the ground state.

**INTRODUCTION**

During the state of equilibrium, proteins in solution fluctuate over a broad spectrum of structural states, ranging from folded to unfolded conformations (1–3). Structural adaptability and the ability to facilitate transitions from the basic folded state (i.e., the ground state) to higher Gibbs free energy states are both necessary for various protein functions, including signal transduction (4) and enzymatic reactions (5). Multiple nuclear magnetic resonance (NMR) techniques, such as $R_2$ dispersion (5, 6) and $^1H/^2H$ exchange NMR spectroscopy (7, 8) have been widely used to study the intrinsic conformational fluctuation in proteins. Our group and several other groups have previously shown that multi-dimensional NMR spectroscopy combined with a pressure perturbation (i.e., high-pressure NMR spectroscopy) is a powerful tool that allows us to study the conformational fluctuations of a protein over a wide spatial and temporal range (9–19). High-pressure NMR spectroscopy has demonstrated that pressure-stabilized states are similar in structure and dynamics to the intrinsically populated states at atmospheric pressure (20, 21), and that high-energy states are evolutionarily conserved among homologous proteins (22, 23). Although the three-dimensional structure, conformational dynamics, controlled thermodynamics, and functions of proteins should be manifested by their particular amino acid sequences, the natural rules for molecular design remain largely unknown.

Internal cavities are important structural elements that produce conformational fluctuations in proteins. We have recently shown that cavities are conserved at similar positions in lysozymes derived from a variety of biological species, and that they are responsible for producing certain types of selective disorder in proteins (24). Moreover, from a thermodynamic viewpoint, it was demonstrated that cavity volume in a folded protein represents a major contribution on the volume changes upon protein unfolding (25). In order to gain further insights into the role of internal cavities of proteins, we investigated the conformational fluctuations in the cavity mutant L99A of T4 lysozyme, using high-pressure NMR spectroscopy. In particular, we focused our attention on the NMR resonances from the methyl groups that line the hydrophobic cavities of this protein.

Wild-type T4 lysozyme is a small hydrolytic enzyme, comprising two domains (N-terminal domain and C-terminal domain), connected by a long central helix, designated as the 'C-helix' (Fig. S1*a* in the Supporting Material). The protein contains hydrophobic and hydrophilic cavities. An L99A (Leu-99 to Ala-99) mutation increases the size of the pre-existing hydrophobic cavity to approximately 150 Å$^3$ in the core of the C-terminal region (26) (Fig. S1*b* in the Supporting Material). The L99A mutant has been used as a model system for understanding protein dynamics in the ligand binding process, as the enlarged cavity allows the ligand binding of substituted benzenes (27) as well as xenon (28). Many physicochemical studies and structure determinations have been performed using X-ray crystallography and NMR spectroscopy (6, 27-32). X-ray crystallography showed that the designed hydrophobic cavity in the L99A mutant is sterically inaccessible to the incoming ligands, yet the protein is able to rapidly bind to benzene and to other similar ligands in solution (29). In contrast, the $R_2$ dispersion-based NMR analysis of L99A identified a slow timescale motion around the cavity, which is a conformational transition between the ground state (native state) and an excited state (i.e. high free-energy state) (6, 30, 31). Structural modeling based on chemical shifts (CS-Rosetta) identified a conformation for the transiently formed state of the protein, in which the aromatic side chain of phenylalanine 114 (F114) was flipped into the enlarged cavity with a simultaneous reorientation of the F-helix (32) (Fig. S1*c* in the Supporting Material). Similarly, in the case of the L121A/L133A cavity mutant of T4 lysozyme, enlargement of the cavity led to a more prominent population of an alternatively packed conformation, in which the cavity was filled, or partially filled, with an engineered side-chain (i.e. nitroxide) (33). In this paper we show that the application of high pressure to T4 lysozyme L99A leads to a redistribution of enzymes in the stable conformation, resulting in a higher occupancy of enzymes in the excited state.

**MATERIALS AND METHODS**

**NMR sample preparations**

Recombinant cysteine-free T4 lysozyme L99A as a cavity mutant of pseudo wild-type (C54T/C97A; WT*) was expressed and purified by the methods described previously (34). Uniformly labeled $^{15}$N- and $^{13}$C-T4 lysozyme was expressed in the M9 media with $^{15}$NH$_4$Cl and $^{13}$C-glucose as the sole sources of nitrogen and carbon, respectively. The protein sample for the NMR measurements was dissolved in 50 mM phosphate buffer (pH 6.0, 10% $^2$H$_2$O), containing 25 mM sodium chloride at a concentration of 1 mM. DSS (2, 2-dimethyl-2-silapentane-5-sulfonate) was used as an internal reference for the $^1$H chemical shifts.

**High-pressure NMR measurements and simulation of the NMR spectra**

High-pressure NMR experiments were performed using a home-made pressure-resistive quartz cell with an outer diameter of ≈ 3.5 mm and an inner diameter of ≈ 1 mm in the pressure range 3–300 MPa, on a DRX 800 spectrometer (Bruker Biospin Co.) (9). To avoid bubbling, we maintained the lowest pressure at 3 MPa, instead of at 0.1 MPa. The one-dimensional NMR experiments were performed at a $^1$H frequency of 800.16 MHz, using a 3-9-19 pulsed field gradient for water suppression. In addition, the $^1$H/$^{13}$C heteronuclear single quantum correlation (HSQC) measurements were recorded at a proton frequency of 800.16 MHz and a $^{13}$C frequency of 201.20 MHz, with 128 increments. At all pressures, the $^1$H chemical shifts were referenced to the methyl signal of DSS, and the $^{13}$C chemical shift was indirectly referenced to DSS (0 ppm for $^1$H). The data points were extended to 2048 × 256, and 90 degree shifted sine-bell window functions were applied to each dimension. All the data were processed using the programs NMRPipe (35) and NMRview (36). Crosspeak intensities (volumes) and chemical shifts were obtained by NMRview. When crosspeaks overlap seriously, the intensity data are not analyzed.

The simulations of the $^{13}$C NMR spectra were performed using WINDNMR-Pro (37). The NMR spectra were simulated as a function of the population of the high-energy state, assuming a two-state exchange model. The exchange rate constant, $k_{ex}$, and the chemical shift difference ($\Delta\omega$) between the ground- and high-energy states of T4 lysozyme L99A were used in the simulation (31).

**3D-RISM calculations**

The three-dimensional reference interaction site model (3D-RISM) is a computational solvent representation that is based on the statistical liquid theory (38,39) (see the Supporting Material). 3D-RISM and the prerequisite 1D-RISM solvent susceptibility calculations were performed using AmberTools 12 (40). The solvent was prepared as pure water using the SPC/e water model (with modified hydrogen radii), using 1D-RISM within the *rism1d* program. The solvent density was modified to meet the experimental water densities at each pressure. Three solutes (the ambient pressure

X-ray, high pressure X-ray, as well as the excited state NMR modeled structures of the T4 lysozyme L99A) are 2B6Y, 2B6X, and 2LCB, respectively, in the Protein Data Bank. The solvent was completely removed, and the protonation states were set at a pH of 7.0, using PROPKA (41-43) in the PDB2PQR program (44,45). The NMR modeled structure included two more C-terminal residues (Asn and Leu), which were removed in order to allow a fair comparison. An optimized multiple-minimization approach was used to find the lowest local free energy structure (generalized born solvent with conjugate gradient minimization, generalized born solvent with L-BFGS minimization, and 3D-RISM with L-BFGS minimization) using the *nab* program in AmberTools 12. All the reported data were obtained from the final 3D-RISM distribution functions for the free-energy minimized structures.

Cavity occupancies were calculated at the appropriate pressure by integrating the population function, $P$, within a sphere encapsulating the cavity,

$$P = \rho \int_{V(cavity)} g_O(\vec{r}) \cdot d\vec{r} \quad (1),$$

where $\rho$ is the average density of solvent species, $g_O$ is the 3D-RISM distribution function of water oxygen (see the Supporting Material). Excess translational entropies were taken into account by simply integrating the expression for the solvent-water excess translational entropy, $S$, over the appropriate volume (46),

$$S = -\frac{1}{2} \rho k_B \int_{V(cavity)} g_O(\vec{r}) \cdot \ln g_O(\vec{r}) \cdot d\vec{r} \quad (2).$$

## RESULTS AND DISCUSSION

**Pressure-induced chemical shift changes**

Recombinant cysteine-free T4 lysozyme (C54T/C97A) (pseudo wild-type; WT*) and its cavity mutant L99A (C54T/C97A/L99A) were created previously to study the role of the hydrophobic cavity of the protein (26). The basic folded conformation (the ground state) of L99A obtained by X-ray crystallography was closely similar to that of the WT*, as shown in Fig. S1 in the Supporting Material.

$^1$H one-dimensional NMR and two-dimensional $^1$H/$^{13}$C HSQC spectra of uniformly

double-labeled ($^{13}$C/$^{15}$N) L99A were collected at different pressures up to 300 MPa. One-dimensional $^1$H NMR spectra at various pressures from 3 to 300 MPa indicated that the protein maintains its folded conformation even at 300 MPa (Fig. S2 in the Supporting Material). Figure 1*a* shows a superposition of the $^1$H/$^{13}$C HSQC spectra at different pressures, up to 300 MPa. The chemical shift changes and the signal intensities are observed as a function of increasing pressure, and the spectral changes are fully reversible. The $^1$H and $^{13}$C chemical shifts varied linearly with pressure, up to a pressure of 100 MPa. Figure 1*b* shows the chemical shift changes of the side-chain methyl groups at a pressure of 100 MPa ($^1$H$^C$, $^{13}$C$^H$). The averages and the root mean-square deviations (R.M.S.D.) of the pressure-induced shifts in $^1$H$^C$ and $^{13}$C$^H$ were -0.029 ± 0.017 ppm and 0.19 ± 0.14 ppm, respectively. These linear pressure-induced shifts generally occur because of a small linear compression of the tertiary protein structure, in the ground state ensemble of the protein (47). However, pressure-induced chemical shifts were largely different at different sites (Fig. 1*b*). Below 100 MPa, the methyl groups showing large deviations from the average behavior were located around the enlarged cavity for L99A (e.g., I78γ$_2$, L79δ$_2$, L84δ$_2$, V94γ$_1$, V94γ$_2$, A98β, I100γ, M102ε, V103γ$_1$ V103γ$_2$, M106ε on the C- to E-helices, L121δ$_2$ on the G-helix, L133δ$_2$ and A134β on the H-helix, and I150δ$_1$ as well as I150γ$_2$ on the J-helix), as indicated by the green spheres in Fig. 1*c*. These results indicate that mechanical compression is much more significant around the cavities than in the rest of the protein.

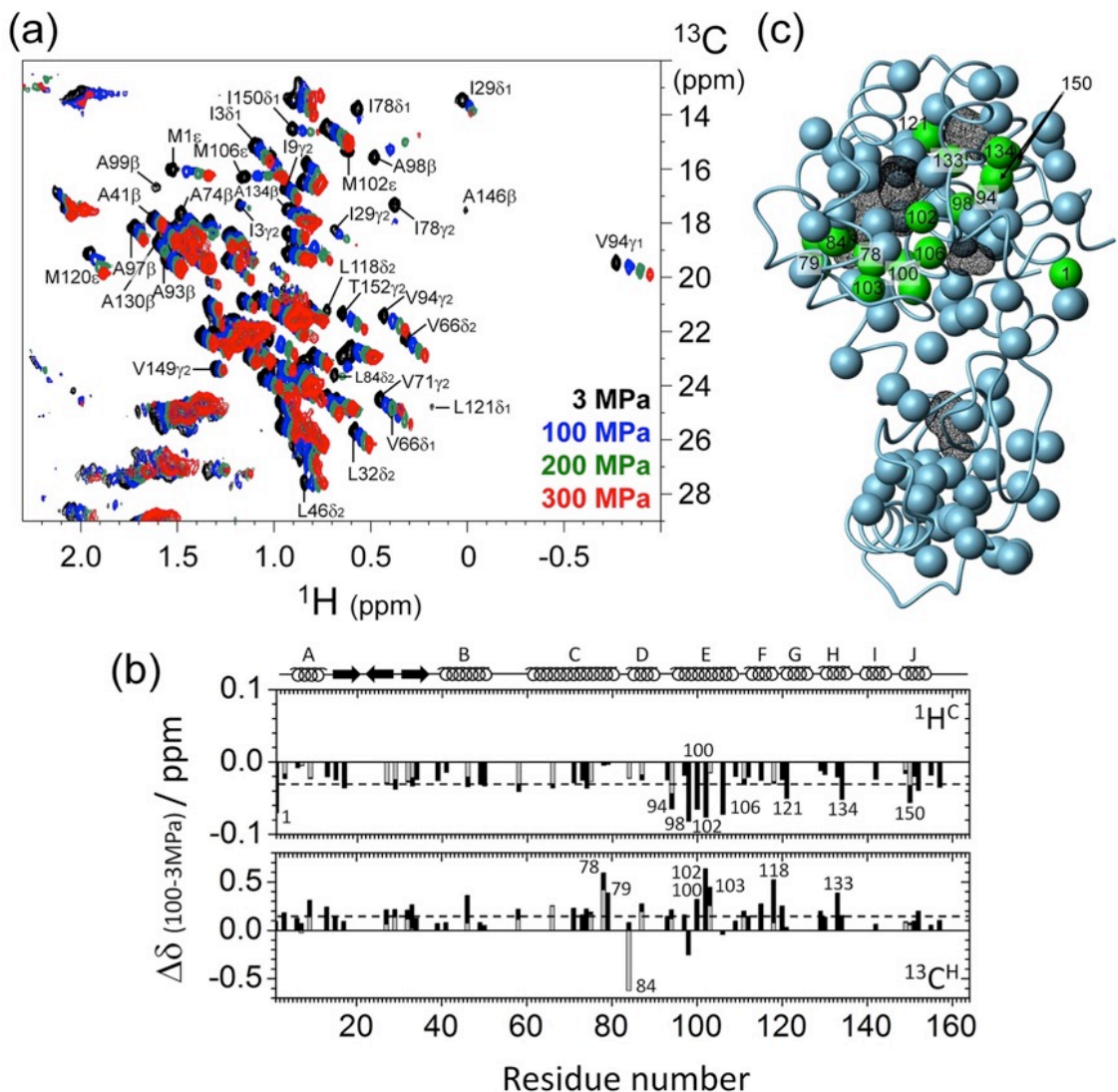

**FIGURE 1.** Effects of pressure on L99A T4 lysozyme. (*a*) $^1$H/$^{13}$C HSQC spectra at different pressures: 3 MPa (*black*), 100 MPa (*blue*), 200 MPa (*green*), and 300 MPa (*red*) for a 1 mM uniformly $^{13}$C- and $^{15}$N-labeled protein at 25°C in 50 mM phosphate buffer (pH 6.0, 10% $^2$H$_2$O/90% $^1$H$_2$O mixture (*v/v*)) containing 25 mM NaCl. (*b*) Pressure-induced chemical shift changes in the low-pressure region ($\Delta\delta = \delta_{100\text{MPa}} - \delta_{3\text{MPa}}$). Chemical shift changes of the methyl proton $^1$H$^C$ and the methyl carbon $^{13}$C$^H$ are plotted along with the residue number. The average values of the pressure shift are -0.03 for $^1$H$^C$ and 0.19 for $^{13}$C$^H$, as shown by the dotted lines. Residues showing large pressure-induced chemical shift changes are represented by the residue numbers. The α-helix and β-strand regions are indicated by rings and arrows, respectively, at the top of the panel. (*c*) Methyl groups showing large pressure-induced shifts ($\Delta\delta > |0.05|$ ppm for $^1$H$^C$ and/or $\Delta\delta > |0.3|$ ppm for $^{13}$C$^H$) are depicted by green spheres on the tertiary

structure of the L99A mutant (PDB ID: 1L90), whereas the less pressure-sensitive methyl groups are depicted by grey spheres. The internal cavities are drawn with a black wire frame cage calculated using a 1.4 Å probe and the program MOLMOL (54).

---

**Pressure-induced changes in the signal intensity**

Figure 2*a* shows changes in peak intensities (i.e. maximum peak height) of the $^1$H/$^{13}$C HSQC spectra of L99A, as a function of pressure. Errors in intensity estimation are shown in Fig. S3 in the Supporting Material. Several crosspeaks of the side-chain methyl groups displayed a reduced intensity with increasing pressure, up to 200 MPa. Above 200 MPa, several of the other crosspeaks also decreased in intensity, however, new signals corresponding to a disordered polypeptide chain were not observed in the HSQC spectra (Fig. 1*a*). The change in peak intensity, namely a broadening or absence of crosspeaks, presumably indicates a heterogeneity of conformations and/or fluctuation among the conformations within the NMR chemical shift time scale (≈ ms). The loss of total crosspeak volumes (i.e. integrated peak intensity) in the HSQC at 300 MPa equated to a 26% reduction of the original volumes (the sum of the integrated crosspeak intensities in Fig. 2*a*). In the case of conventional $^1$H/$^{13}$C HSQC, increases in $^{13}$C spin-spin relaxation in the $t_1$ evolution period (i.e. 0-9.6 ms), as well as $^1$H spin-spin relaxation in $t_2$ period (i.e. 98.4 ms) are likely responsible for the observed crosspeak broadening and loss of crosspeak intensities. In addition, transfer amplitudes of the two INEPT steps, in total 7.5 ms, may decrease with pressure if $^1$H spin-spin relaxation increases.

The crosspeaks of the methyl groups can be classified into three groups: (i) a rapidly decaying group (red lines), (ii) an intermediate decaying group (yellow lines), and (iii) a slowly decaying or unaffected group (black lines). The methyl groups belonging to classes (i) and (ii) are shown as red and yellow spheres, respectively, in the structure of L99A (Fig. 2*b*). All of the rapidly decaying signals from group (i) (I78$\delta_1$, I78$\gamma_2$, L84$\delta_1$, A99$\beta$, M102$\epsilon$, L118$\delta_2$, L121$\delta_1$, and L133$\delta_2$) involved the methyl groups lining the enlarged hydrophobic cavity. Members of the intermediate group (ii) (I3$\delta_1$, I3$\gamma_2$, M6$\epsilon$, L7$\delta_1$, I29$\delta_1$, I29$\gamma_2$, L46$\delta_1$, L66$\delta_1$, V71$\gamma_2$, V75$\gamma_2$, L84$\delta_2$, V87$\gamma_2$, A98$\beta$, I100$\gamma_2$, M106$\epsilon$, A129$\beta$, A130$\beta$, I150$\delta_1$, and I150$\gamma_2$) were located around the enlarged cavity, as well as other cavities of the protein (Fig. 2*b*).

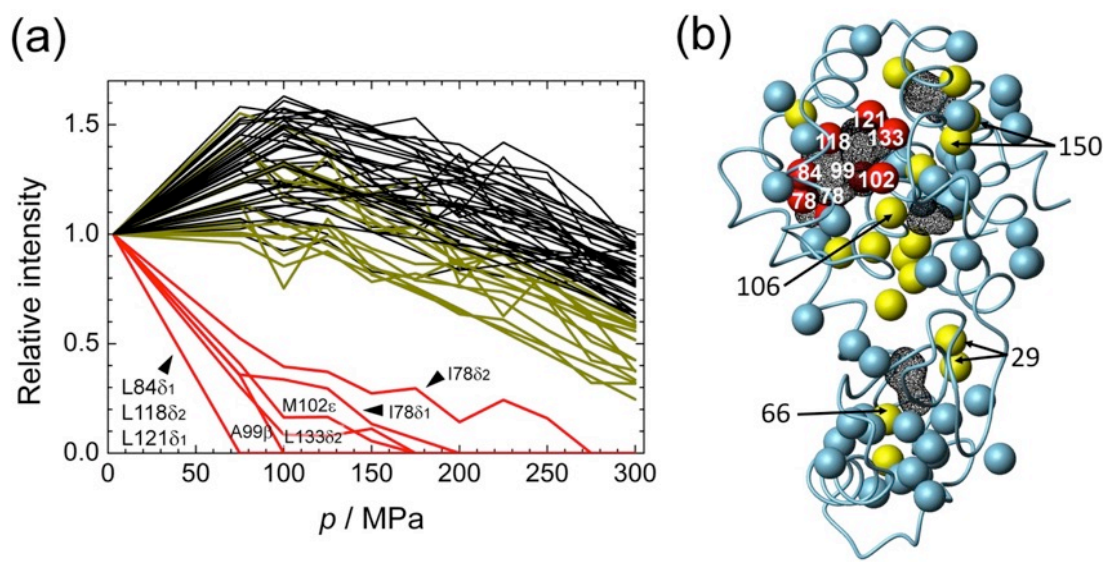

**FIGURE 2.** (*a*) Plots of maximum height (normalized at 3 MPa) of the crosspeaks in $^1$H/$^{13}$C HSQC for L99A, obtained at various pressures, from 3 to 300 MPa. Changes in the protein concentration as a function of pressure are corrected, based on the pressure-induced compaction of the solvent water (by ≈ 9% at 300 MPa) (63). The crosspeaks of the side-chain methyl groups can be classified mainly into three groups: (i) the rapidly decaying group (*red lines*), (ii) the intermediate decaying group (*yellow lines*), and (iii) the slow decaying group (*black lines*). The rapid and the intermediate decaying groups comprise (i) I78$\delta_1$, I78$\gamma_2$, L84$\delta_1$, A99$\beta$, M102$\epsilon$, L118$\delta_2$, L121$\delta_1$, and L133$\delta_2$ and (ii) I3$\delta_1$, I3$\gamma_2$, M6$\epsilon$, L7$\delta_1$, I29$\delta_1$, I29$\gamma_2$, L46$\delta_1$, L66$\delta_1$, V71$\gamma_2$, V75$\gamma_2$, L84$\delta_2$, V87$\gamma_2$, A98$\beta$, I100$\gamma_2$, M106$\epsilon$, A129$\beta$, A130$\beta$, I150$\delta_1$, and I150$\gamma_2$, respectively. Errors in height estimation are presented in Fig. S3. (*b*) The observable side-chain methyl groups in the HSQC spectrum at 3 MPa (Fig. 1*a*) are depicted with spheres. Those belonging to the rapid and intermediate decaying groups are represented by red and yellow spheres, respectively, on the tertiary structure, and residues located around the internal cavities are designated by their residue number. The picture was prepared using MOLMOL (54).

---

**Conformational fluctuations within the ground state ensemble**

The partial molar volume of a protein fluctuates with thermal agitation in two different ways: A fluctuation either occurs within the same sub-ensemble of conformers, or occurs because of a transition into a different sub-ensemble of conformers possessing a higher Gibbs free energy (referred to as an excited state). Thus, a protein biomolecule adopts a smaller partial molar volume using either or both of the aforementioned adaptations.

Fluctuations occurring within the same sub-ensemble of conformers can be studied by observing the pressure-induced changes in the chemical shifts and peak intensities. The $^{13}$C chemical shifts reflect site-specific structural changes at the side chains of the protein, whereas the amide $^{1}$H and $^{15}$N chemical shifts generally indicate the response of the hydrogen-bond distance and torsion angles to pressure (47). Furthermore, the methyl $^{13}$C chemical shifts are more sensitive to atomic packing. Regardless of the underlying dynamics, any change in the measured chemical shifts corresponds to the average of all the structural changes occurring rapidly on the NMR chemical shift time scale ($\approx$ ms). We interpret the linear chemical-shift change observed in the lower pressure region (up to $\approx$ 100 MPa) to be the result of a linear change in the average inter-nuclear distances and torsion angles (11), the combination of which determines the compressibility coefficient ($\beta_T$) of the protein.

Larger deviations from the average pressure-induced shifts are observed for the $^{13}$C nuclei of methyl groups lining internal cavities, particularly those at the enlarged cavity of L99A. These variations indicated that larger structural changes, including the compression of inter-atomic distances, take place around the cavities, rather than elsewhere inside the protein structure. As the macroscopic compressibility (which is the net sum of the microscopic compressibilities) is statistically correlated with the volume fluctuation of the system consisting of the solute and the solvent (48), the larger volume fluctuations occurring rapidly on the chemical shift time scale could be attributed to the enlarged cavity of the protein. Similar large pressure-induced changes in the chemical shifts were also observed around the internal cavities of BPTI (49) and hen egg white lysozyme (50). Our past and present results both indicate that volume fluctuations are greater at the residues closer to the internal cavities (at atmospheric pressure), as these regions presumably have a larger structural adaptability to pressure.

In the same pressure range as above (up to 100 MPa), the crosspeaks of the side-chain methyl moieties involved in the rapidly decaying group (i) (e.g., I78$\delta_1$, I78$\gamma_2$,

L84δ$_1$, A99β, M102ε, L118δ$_2$, L121δ$_1$, and L133δ$_2$) significantly lose their intensities with increasing pressure. The losses in peak intensities in two-dimensional HSQC spectra at varying external parameters, such as pressure and temperature, are generally attributed to an increase in peak-widths (i.e. spin-spin relaxation rates, $R_2$, of the nuclei) in both dimensions, which might result from heterogeneous conformations within the basic folded ensemble, or from conformational exchange among different stable conformational states of the protein on the microsecond to millisecond time scale. As discussed below, the significant decrease in the peak intensities for some of the methyl groups cannot be explained only with the conformational exchange between the ground state and the particular high-energy state of the protein. It is presumably caused by heterogeneous configurations of the corresponding methyl groups resulting from the compression and modified hydration of the enlarged cavity.

In contrast, we observed increases in the crosspeak intensities for many of the intermediate and slow decaying methyl groups, up to 100 MPa. The same was observed for signals of the backbone in the $^1$H/$^{15}$N HSQC data (not shown) for L99A. Importantly, we did not observe this effect for WT*. At the moment we have no explanation for this effect, other than the possibility that it results from the formation of the cavity in the C-terminal domain.

**The effect of increasing pressure on the high-energy excited state of L99A**

Recently, the structure of the transiently formed high-energy state of T4 lysozyme L99A was characterized by employing a combined strategy comprising $R_2$ dispersion NMR and CS-Rosetta model building (32). Clear structural differences between the ground- and high-energy states were observed, particularly in the vicinity of the enlarged cavity of the protein. In the high-energy state, the side chain of F114 was accommodated inside the cavity, together with a simultaneous reorientation of the F-helix, which required fluctuations outside the ground state ensemble of the protein. $^{15}$N-$R_2$ dispersion NMR experiments showed that the high-energy state was populated to about 3% at 25°C and atmospheric pressure, and was in equilibrium with the ground state (97% populated). The corresponding interconversion occurred on the millisecond timescale, leading to pervasive resonance broadening.

In the present high-pressure NMR study, we observed severe line-broadening and missing HSQC crosspeaks, particularly for the residues around the enlarged cavity, at pressures below 200 MPa. Fig. 3 shows a representation of the methyl groups exhibiting a relatively large chemical shift difference ($\Delta\omega$) for the $^{13}$C nuclei (Fig. 3*a*) between the ground and high-energy states, based on the $^{13}$C-$R_2$ dispersion measurements (31), along with the resonance showing severe line-broadening with increasing pressure (Fig. 3*b*). Interestingly, the latter group of residues matches closely with the group showing larger chemical shift changes upon conformational transition to the transiently formed high-energy state at atmospheric pressure.

Unfortunately, similar data for the methyl protons of T4 lysozyme L99A are not available. However, the closely similar mutant L99A/G113A was studied previously, and was shown to undergo a very similar exchange between ground and excited states, albeit at a slower rate. We were thus able to use the $^1$H $\Delta\omega$ values of L99A/G113A, which were measured by ZZ-exchange NMR at 1 °C (32). Exchange crosspeaks in the ZZ-exchange NMR spectrum were observed for at least 13 methyl groups (including I3H$\gamma_2$, I27H$\gamma_2$, I29H$\delta_1$, L66H$\delta_1$, V71H$\gamma_2$, A74H$\beta$, I78H$\gamma_2$, A98H$\beta$, A99H$\beta$, M102H$\varepsilon$, I118H$\delta_2$, L121H$\delta_1$, and A146H$\beta$). Fig. 3*c* shows the distribution of these methyl groups based on the chemical shift difference ($\Delta\omega$) for the $^1$H nuclei. The group showing severe line-broadening with increasing pressure again matches closely with the group showing larger $\Delta\omega$ for the $^1$H nuclei.

These imply that the broadening of the HSQC crosspeaks with increased pressure, and the corresponding increase in the exchange contribution to the $^{13}$C and $^1$H transverse spin relaxation rates, can likely be explained by an increase in the high-energy state population and/or a decrease in the exchange rate constant $k_{ex}$, at high pressure.

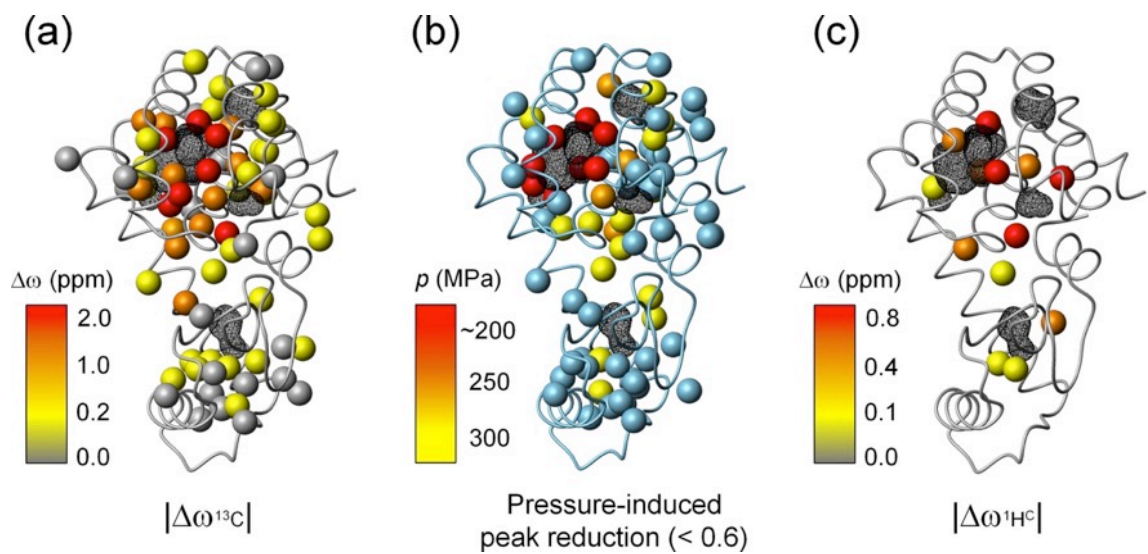

**FIGURE 3.** Distribution of residues based on (*a*) $^{13}$C chemical shift differences ($|\Delta\omega|$) for side-chain methyl group between exchanging sites, as estimated by the $R_2$ dispersion NMR experiments for L99A (31). The amplitude of $|\Delta\omega|$ is represented by a color scale from yellow to red. (*b*) Side-chain methyl groups showing significant loss of their crosspeak intensities at a high pressure. The pressure at which the relative peak intensity is reduced to less than 0.6 from the original is represented by colors: <200 MPa; *red*, 250 MPa; *orange*, and 300 MPa; *yellow*. (*c*) Distribution of residues based on the $^1$H chemical shift differences ($|\Delta\omega|$) for side-chain methyl groups between exchanging sites, as estimated by the ZZ-exchange NMR experiments for L99A/G113A (32). The picture was prepared using MOLMOL (54).

---

**Line-shape simulation**

In order to test the hypothesis that the broadening of the HSQC crosspeak intensities with increased pressure can be explained by an increase in the high-energy state population and/or a decrease in the exchange rate constant $k_{ex}$, we simulated the changes in the $^{13}$C and $^1$H spectra using the WINDNMR-Pro program (37). The $k_{ex}$ and $^{13}$C chemical shift difference ($\Delta\omega$) between the ground- and the transiently formed high-energy states of L99A are available in the literature (31). Since $\Delta\omega$ values for the methyl protons of L99A have not been reported, we used the $\Delta\omega$ values for the L99A/G113A mutant of T4 lysozyme (32), which were measured by ZZ-exchange NMR at 1 °C.

Although, in reality, both the population of the high-energy state and the exchange rate constant $k_{ex}$ could change simultaneously with increasing pressure, we first simulated $^{13}$C spectral changes as a function of the population of the transiently formed high-energy state. Fig. 4a shows the simulated $^{13}$C spectra for L121C$\delta_1$, which is a representative carbon nucleus showing a rapid decrease in the HSQC peak intensity as a function of pressure (group (i)). When the global exchange rate is 1449 s$^{-1}$, $\Delta\omega$ is 1.88 ppm (1.88×201.2×2π rad s$^{-1}$) (31), which corresponds to a slow-to-intermediate exchange condition ($k_{ex}/\Delta\omega$ = 0.61). Fig. 4b shows a simulation for L84C$\delta_2$, which belongs to the intermediate decaying group (ii). Its $\Delta\omega$ value is 0.37 ppm (0.37×201.2×2π rad s$^{-1}$) (31), which correlates with an intermediate-to-fast exchange condition ($k_{ex}/\Delta\omega$ = 3.1). The lineshape simulations clearly show that for both of the residues mentioned above, the peak-width of the ground state rapidly increases with increasing high-energy state population, and that the peak almost disappears when the populations become similar. A new peak, corresponding to the high-energy state, then starts to appear. Other examples of groups that decay at a rapid or intermediate pace are presented in Fig. S4 in the Supporting Material.

Secondly, we simulated the $^{13}$C spectra at variable $k_{ex}$ values, assuming the population of the high-energy state to be constant (i.e. 3.4%) (Fig. 4c and d). Changes in peak-widths corresponding to the ground and high-energy states are not substantial as long as the high-energy state population is low. However, the peak-widths may show larger changes with increased high-energy state populations. The $k_{ex}$ usually decrease under high pressure because the positive activation volume for folding is larger than negative activation volume for unfolding (51-53). Thus, in the case of both slow and intermediate exchange conditions, peak-widths would be expected to be sharper under high pressure.

Thirdly, we simulated $^{1}$H spectra as a function of the high-energy state population and varying $k_{ex}$ values (Fig. 4e and f). In the case of L121 H$\delta_1$, the $\Delta\omega$ was 0.43 ppm (0.43×800.13×2π rad s$^{-1}$), which corresponds to a slow-to-intermediate exchange condition ($k_{ex}/\Delta\omega$=0.68). When the population of the high-energy state rises, the $^{1}$H peak width rapidly increases in a manner similar to the $^{13}$C peak-width (Fig. 4a). (see Fig. S5 for other examples of methyl protons involved in the rapid and intermediately decaying groups). Therefore, peak-broadening in $^{1}$H/$^{13}$C HSQC spectra would occur due to both $^{1}$H and $^{13}$C spin relaxation, dependent on the chemical shift difference ($\Delta\omega$). These $^{13}$C and $^{1}$H spectral simulations strongly suggest that an increase in the high-energy state population is responsible for the selective peak-broadening observed with increasing pressure. In the case of the conventional HSQC, peak-broadening would occur during the $t_1$-evolution period (i.e. 0-9.6 ms) due to $^{13}$C spin-spin relaxation, and during the $t_2$-ditection period (i.e. 98.4 ms) due to $^{1}$H spin-spin relaxation. In addition, transfer amplitudes of the two INEPT steps (i.e. 7.5 ms) may decrease with pressure if $^{1}$H spin-spin relaxation increases. Because dipolar relaxation on spin-spin relaxation rates ($R_2$) may not alter much with pressure, exchange contribution on $R_2$ of both nuclei could be responsible for the broadening and missing of the signals.

We noticed that signals corresponding to the high-energy state are not observed in the spectra under high pressure. Two explanations for this could be that, in general, a population of the minor state is low and NMR peaks of the minor state have larger peak-widths than those of the major state (see Fig. 4), and thus, they exhibit a relatively lower signal-to-noise ratio. It is also possible that the high-energy state corresponds to much larger peak-widths than the ground state because of conformational heterogeneity; note that intrinsic peak-width for the high-energy state was assumed to be same as that for the ground state in the spectral simulation.

For some of the signals in the intermediate decaying groups, the peaks showed sigmoidal chemical shift changes, as well as intensity losses, as a function of the high-energy state population (M6ε, I29γ$_2$, and A130β, Fig. S4*b* in the Supporting Material). It is not surprising that, as a function of the high-energy state population, the chemical shifts and the line broadening, depend strongly on the magnitudes of $k_{ex}$ and $\Delta\omega$. Our simulations show that the line broadening caused by pressure changes could be explained by an increase in the high-energy state population. A substantial correlation between the line broadening (large $R_2$) at ambient pressure and the peak-intensity reductions (for the same probes) at higher pressures strongly suggests that both methods are probing the same conformational equilibrium, albeit with different populations of the associated states.

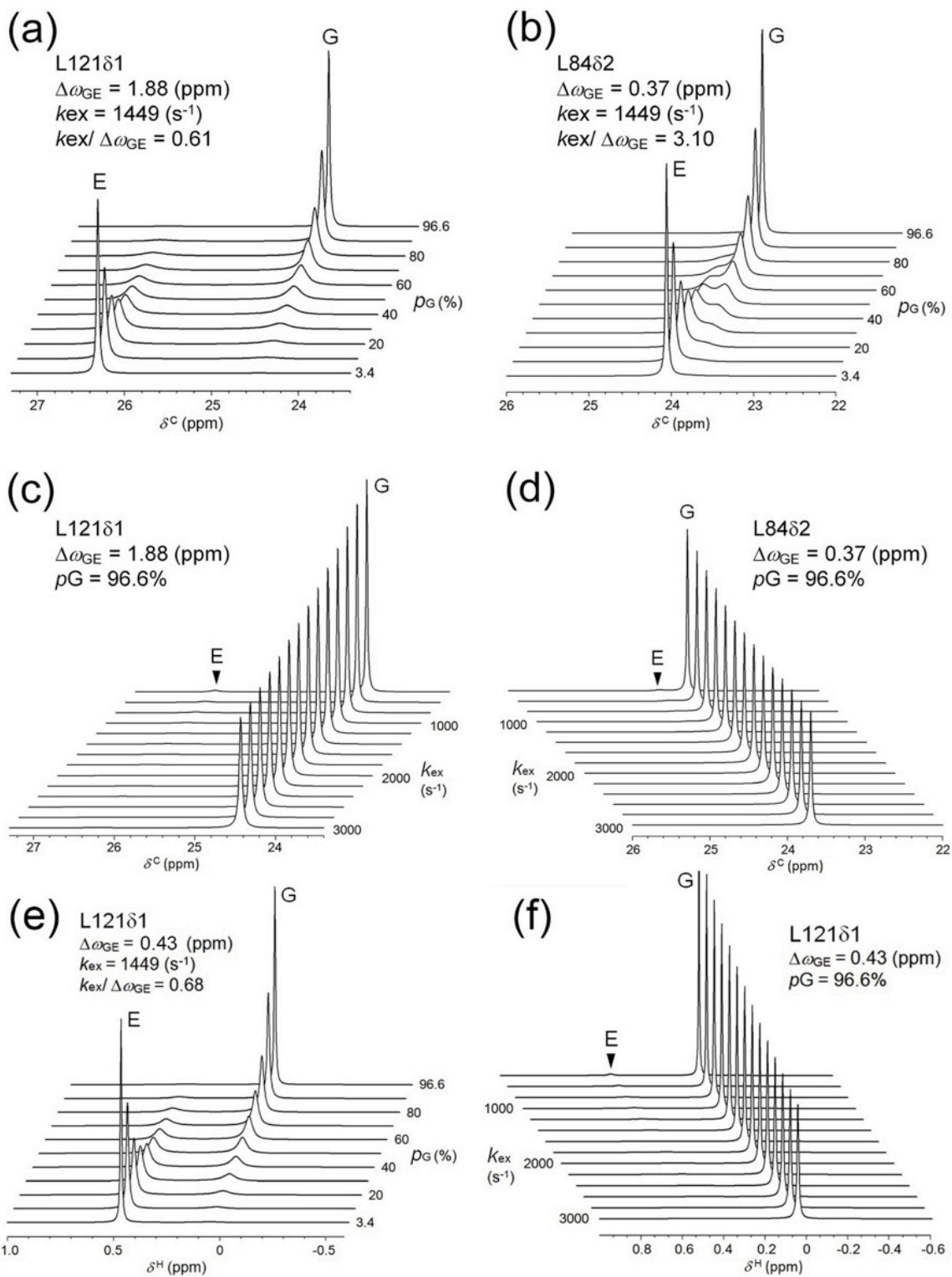

**FIGURE 4.** Line-shape simulation for the methyl carbon and proton peaks of L99A, in accordance with the two-state exchange model. (*a-b*) $^{13}$C line-shape spectral simulation for the L121C$\delta_1$ carbon (*a*) and for the L84C$\delta_2$ carbon (*b*), as a function of the excited state population. (*c-d*) $^{13}$C line-shape spectral simulation for the L121C$\delta_1$ carbon (*c*) and the L84Cδ$_2$ carbon (*d*), as a function of $k_{ex}$. The $^{13}$C chemical shift difference Δω$_{GE}$ between the ground (G) and the transiently populated high-energy state (E), the rate constant $k_{ex}$ (1449 s$^{-1}$) for a chemical exchange between the two states, and a population of the high-energy state $p_E$ (3.4%) were all obtained from the literature (31). (*e-f*) $^1$H line shape spectral simulation for L121Hδ$_1$ proton, as a function of the high-energy state population (*e*), and as a function of $k_{ex}$ (*f*). The $^1$H chemical shift values of the ground and high-energy states of L99A/G113A, measured by ZZ-exchange NMR experiments at 1 °C (32), are used in the simulations. The program WINDNMR-Pro was used for the simulations (37).

---

Next, we consider the differences in the peak intensity reductions between the intermediately and rapidly decaying groups. When an increase in line-width (i.e. $R_2$) occurs in both $^1$H and $^{13}$C dimensions, peak intensities may decrease rapidly with an increase in the high-energy state population. This applies to L121δ$_1$ (the rapidly decaying group), but not to I3γ$_2$ (intermediate decaying group), although each of these has relatively large Δω values for both the $^{13}$C and the $^1$H chemical shifts. We did not find clear differences in the peak intensity reductions between the intermediately and rapidly decaying groups. In other words, the significant loss in the peak intensities for the rapidly decaying methyl groups could not be explained solely based on the conformational exchange between the ground state and the particular high-energy state of the protein.

As a comparison, we studied pressure effects on the WT* protein, and found that rapid signal losses were observed for the methyl groups of M102ε, L121δ$_1$, and L133δ$_2$ (data not shown), all of which are located around the same (but smaller) cavity in the C-terminal region. The rapid signal loss in the WT* protein cannot be attributed to the transition into the high-energy state, because the volume of the cavity in the WT* protein is too small (39 Å$^3$) to bind the F114 side-chain. Indeed, $^{15}$N spin-spin relaxation NMR analysis did not detect a high-energy state for the WT* protein (30). Therefore, we conclude that the rapid loss of the peak intensities at lower pressure observed for L99A should be attributed to changes in the native state sub-ensemble, namely heterogeneous configurations of the corresponding methyl groups.

**Origins of chemical shift changes between the ground and high-energy states**

Next, we tried to delineate the origins of chemical shift changes by comparing the structures of the ground state (Fig. 5*a* left panel) and the high-energy state (Fig. 5*a* right panel) of the protein. According to the high-energy state structure of L99A (modeled using CS-Rosetta), the F114 aromatic ring is accommodated in the enlarged cavity. Therefore, due to the transition, changes in the ring current shifts are expected. Figures 5*b* and 5*c* show the differences in the ring current shifts between the ground- and high-energy states of the $^1$H and $^{13}$C nuclei, respectively, as calculated by the MOLMOL program, which uses the Johnson-Bovey model. (54). The methyl groups showing large changes due to ring current shifts for at least one of the two nuclei are I78$\gamma_2$, L84$\delta_2$, L91$\delta_1$, V94$\gamma_1$, A98$\beta$, A99$\beta$, M102$\varepsilon$, M106$\varepsilon$, T109$\gamma$, V111$\gamma_2$, L118$\delta_2$, and L121$\delta_1$. All of them are located adjacent to F114, W138, F153, and W158 (Fig. 5*a*). It is obvious that larger differences in the ring current shifts between the ground and high-energy states are observed for the groups located in close proximity to the aromatic rings than those located elsewhere in the protein structure. Moreover, the chemical shifts of the methyl protons and the carbons could be used as sensitive probes for detecting the changes in the orientation of the aromatic rings. Interestingly, a substantial loss in the signal was observed, due to an increased pressure, for the methyl groups lying at the cavity (e.g. I78$\delta_1$, I78$\gamma_2$, L84$\delta_1$, A99$\beta$, M102$\varepsilon$, L118$\delta_2$, and L121$\delta_1$ indicated with asterisks in Fig. 5*b*, respectively), where the F114 aromatic ring is expected to flip-in. The corresponding data for all observed residues is shown in Fig. S6 in the Supporting Material. These results strongly support the idea that the high-energy state, in which the F114 aromatic ring flips into the cavity, is indeed stabilized by higher pressures. It is clear that an increase in the population of the high-energy state would result in the increase in exchange contribution on $R_2$ under high pressure, as the high-energy state exhibits different $^{13}$C and/or $^1$H chemical shifts around the enlarged cavity. The accommodation of the F114 side chain in the cavity would confer the high-energy state a smaller partial molar volume than the ground state, owing to improved packing, and because of an alteration in the hydration structure on the molecular surface.

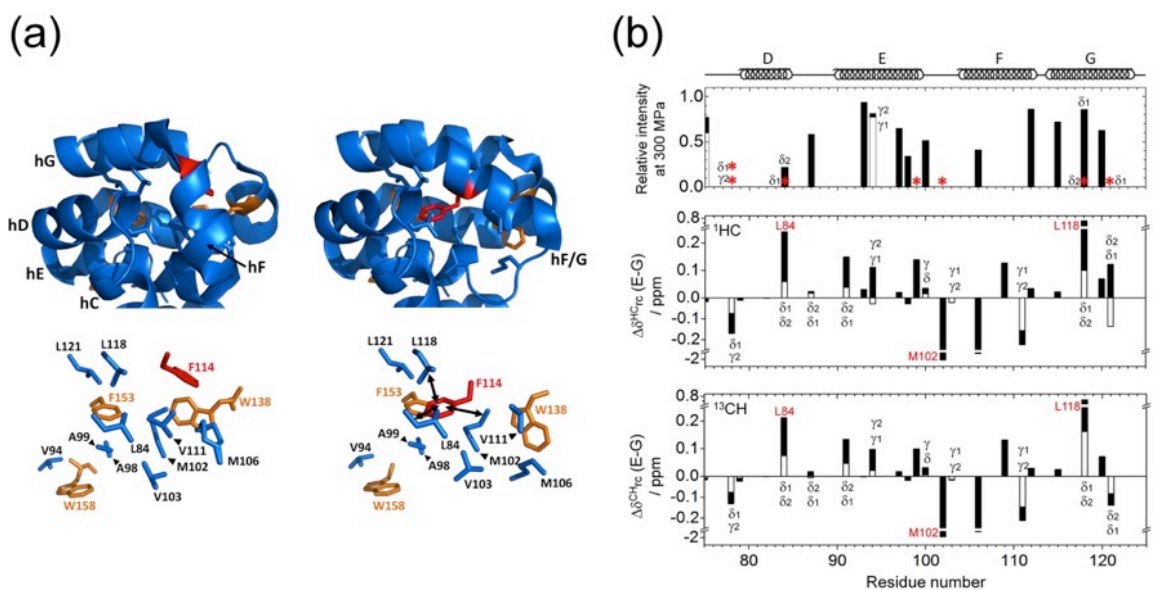

**FIGURE 5.** (*a, top*) A part of the tertiary structure of the ground state (left) and the transiently populated high-energy state (right), as modeled by CS-ROSETTA (32). (*a, bottom*) The hydrophobic side chains around the enlarged cavity are represented by sticks. F114 is colored in red, whereas the other aromatic side chains near the cavity are depicted in yellow. The hydrophobic side chains facing the aromatic rings are shown in blue. (*b, top*) The relative intensities of the $^1$H/$^{13}$C HSQC crosspeaks at 300 MPa (normalized by the values at 3 MPa) for the methyl groups in the vicinity of the enlarged cavity. Methyl groups showing no intensity at 300 MPa are marked with an asterisk. (*b, middle and bottom*) The differences in ring current shifts for the methyl groups between the ground- and high-energy states ($\Delta\delta_{rc}$) for the $^1$H (b, *middle*) and $^{13}$C (b, *bottom*), as estimated by the program MOLMOL (54). The residues facing the F114 side-chain ring in the high-energy state are denoted by red numbers. The attributes for the side-chain methyl groups are denoted with δ and γ, respectively. When the residue has two methyl groups, the group indicated at the bottom shows a larger absolute value than that of the group on the top.

---

Conformational transitions could also be monitored by non-linear responses in the pressure-induced chemical shifts, when $\Delta\omega$ is much smaller than $k_{ex}$ ($\Delta\omega \ll k_{ex}$) (11). Indeed, we observed non-linear responses for the pressure-induced chemical shifts of several methyl groups for L99A above 100 MPa (e.g. for I9, I17, I29, and L79, shown in Fig. S7 in the Supporting Material). These pressure-induced non-linear changes may be a part of the sigmoidal changes seen when the high-energy state population is increased. However, we limited our analysis to the pressure-induced line broadening, as it can be easily monitored in order to delineate the structural motions in the microsecond to millisecond timescale.

**Pressure-induced denaturation**

Above 200 MPa, the intensities of all crosspeaks decreased in a concerted manner (Fig. 2*a*). As predicted by spectral simulation, the peak-width for the ground state increased substantially with the increasing population of the high-energy conformation, even in the slow exchange condition. We therefore suggest that the concerted decrease in the crosspeak intensities above 200 MPa marks the onset of an unfolded conformer 'U', which will not give sharp NMR signals. Because the peaks of the slow decaying group still retain about 80% of their original integrated intensities at 300 MPa, the population of U should account for less than 20% of the conformational ensemble. Keep in mind that NMR peaks of the minor state may not give enough signal to noise ratio to be detected (see Fig. 4). An increase in the unfolded state population is expected under high pressure, since L99A is only marginally stable and the unfolded conformation has smaller partial molar volume than the folded one (9, 25, 55). According to high-pressure fluorescence studies, the unfolded state of L99A should make up about 82% of the population at 300 MPa and 24 °C (e.g. $\Delta G^0$=13.3 kJ/mol, $\Delta V^0$= –56 mL/mol at pH 7.0 and 20 mM NaCl condition) (55). Although this is inconsistent with the present high-pressure NMR data, it is not surprising because tryptophan fluorescence is sensitive to the polarity of its local environment, and the thermodynamic quantities were calculated from a two-state unfolding model. Moreover, because all three tryptophans (W126, W138, and W158) in the protein are located in the C-terminal domain, the pressure-induced changes in fluorescence quenching may report both on transitions into the excited folded and unfolded states.

**Water occupancy in the hydrophobic cavity predicted from 3D-RISM**

Although X-ray crystallography suggested that the enlarged hydrophobic cavity in L99A mutant was sterically inaccessible to incoming ligands, spin relaxation NMR studies indicated the presence of conformational fluctuations around the cavity. This fluctuation was much more amplified in the L99A mutant than in WT*, in solution

(6, 27–29). It is challenging to understand how the hydration of the cavity correlates with the conformational fluctuation of the protein, and how the conformational fluctuations correlate with the ligand-cavity interaction.

Collins et al. demonstrated an increase in the electron density in the enlarged cavity in the L99A mutant with increasing pressure, using high-pressure X-ray crystallography and molecular dynamics simulations (56). The number of water molecules existing in the cavity was predicted to be zero at 0.1 MPa and was predicted to increases to about two at 200 MPa, while the protein structure remained intact. However, a recent high resolution X-ray crystallographic analysis found out that diffuse electron density equivalent to approximately 1.5 water molecules was present in the cavity of the crystallized protein, even at a pressure of 0.1 MPa (57).

Clearly, it is desirable to have a general and comprehensive method for predicting the water occupancy in protein cavities. Hence, we employed the three-dimensional reference interaction site model (3D-RISM), which describes molecular solvation based on a rigorous statistical liquid theory (38). 3D-RISM is suited for analyzing the water occupancy at molecular surfaces, including the internal cavities in proteins (see Materials and Methods) (58-60). 3D-RISM analysis was carried out for three different structures, *viz.* the X-ray structure at 0.1 MPa (*blue*) (2B6Y), the X-ray structure at 200 MPa (*red*) (2B6X), and the NMR data-based ROSETTA model of the transiently populated high-energy conformer at 0.1 MPa (*green*) (2LCB). Free-energy minimized structures were obtained (for the three structures), using 3D-RISM at pressures, 0.1 MPa, 100 MPa, 200 MPa, and 300 MPa, respectively. The partial molar volumes were calculated for the two X-ray structures (2B6Y and 2B6X) at different pressures. The compressed volumes reached about 1.3% for both the proteins, at 200 MPa (isothermal compressibility, $\beta_T \approx 7$ Mbar$^{-1}$), as shown in Table 1. This is consistent with the observation of the macroscopic compressibility of proteins (4–15 Mbar$^{-1}$), obtained using sound velocity measurements (61). Fig. 6*a-c* show the enlarged cavities for the three structures (*a*: 2B6Y, *b*: 2B6X, *c*: 2LCB) at each experimental pressure, as estimated by the conventional sphere rotation method. Figure 6*d* shows the water occupancy in the enlarged cavity for the X-ray structures at 0.1 MPa (*blue*; 2B6Y), 200 MPa (*red*; 2B6X), and for the NMR model of the transiently populated high-energy conformer, at 0.1 MPa (*green*; 2LCB). The water occupancy was increased from 3.37 at 0.1 MPa (2B6Y) to 3.71 (2B6X) at 200 MPa, as shown in Table 1. These data indicate that water occupancy does not increase substantially with increasing pressure, when the basic folded conformation is maintained at the elevated pressure. These 3D-RISM analyses qualitatively support the recent high-resolution X-ray crystallographic analysis, showing the presence of diffuse water in the enlarged cavity of the L99A protein, even at a pressure of 0.1 MPa. However, the number of water molecules theoretically present in the enlarged cavity was two times larger than that predicted by high-resolution X-ray crystallography (57). The diffuse electron density detected using high-resolution X-ray crystallography may correspond to the lower limit of detection, which in turn depends on the data quality and refinement method.

TABLE 1. 3D-RISM analysis.

| Pressure | Partial molar volume | | Water occupancy | | Excess translational entropy | |
|---|---|---|---|---|---|---|
| | 2B6Y Å$^3$ | 2B6X Å$^3$ | 2B6Y * | 2B6X * | 2B6Y J/mol K | 2B6X J/mol K |
| 0.1 MPa | 21310 | 21272 | 3.37 | 3.41 | -42.8 | -42.9 |
| 100 MPa | 21141 | 21141 | 3.60 | 3.58 | -46.0 | -46.2 |
| 200 MPa | 21062 | 20988 | 3.84 | 3.71 | -49.5 | -48.4 |
| 300 MPa | 20865 | 20824 | 3.79 | 3.82 | -50.1 | -51.1 |

*A number of water molecules equivalent to the water occupancy in the hydrophobic cavity.

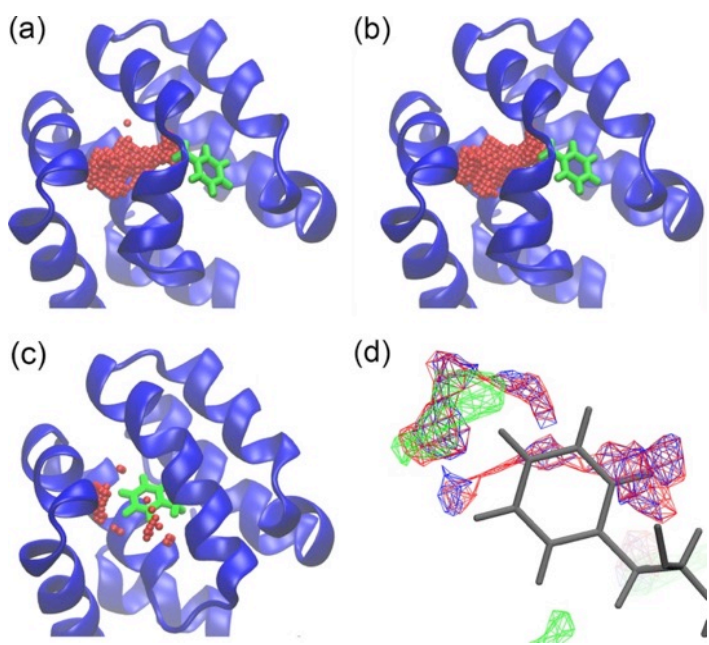

**FIGURE 6.** (*a-c*) A part of the backbone structure of L99A with the enlarged cavities estimated by the conventional sphere rotation method: (*a*) the X-ray structure at 0.1 MPa (2B6Y), (*b*) the X-ray structure at 200 MPa (2B6X), and (*c*) the NMR model of the transiently populated high-energy conformer at 0.1 MPa (2LCB). (*d*) The water occupancy in the enlarged cavity of L99A for these three structures: blue for X-ray structure at 0.1 MPa, red for X-ray structure at 200 MPa, and green for NMR model at 0.1 MPa. The F114 side-chain is shown in stick representation in all panels.

---

Interestingly, the excess translational entropy of the water molecules inside the cavity, as calculated by 3D-RISM, decreased at higher pressures (Table 1, see Materials and Methods). The excess entropy values are negative because they are relative to bulk (bulk excess entropy being zero). The fact that the values get more negative with increasing pressure suggests that, at higher pressures, the molecular positions are more confined, and therefore would be more easily resolved by X-ray crystallography.

As the 3D-RISM analyses described above were carried out only on the energy-minimized structure of the protein at each pressure, they did not take into account the conformational heterogeneity of the protein as seen in solution. As compared to the crystallized protein, the protein in solution may exist as a more varied ensemble of structures, with different levels of hydration. Compression, expansion, and hinge-type motions in proteins at different pressures could bring about changes in the shape and hydration of the cavities in the protein structure. Thus, rotational and diffusive motions of water molecules in the cavities could be altered under high pressure. These dynamic motions of the water molecules and the heterogeneous conformations of the cavities would cause the chemical shifts to disperse. Such an effect could be partly involved in the pressure-induced line broadening of the NMR resonances, observed in the low-pressure region (≈ 100 MPa), as well as in an increase in the population of the high-energy states.

In the case of the transiently populated high-energy state, the aromatic side chain of F114 is flipped into the enlarged cavity, resulting in the displacement of water molecules (Fig. 6*b*). The bulky aromatic ring causes an increase in the atomic occupancy of the enlarged cavity, which may be partly responsible for the small partial molar volume of the high-energy state. Note that the partial molar volume analysis performed using 3D-RISM was not performed on the structural models constructed by CS-Rosetta, as these models have a relatively large uncertainty for the side-chain orientations.

**Cavity as a source of conformational fluctuations**

The present high-pressure NMR experiments and RISM analysis for the T4 lysozyme L99A indicate that the water-containing cavity can serve as a source of conformational fluctuations. Fig. 7 shows a schematic representation of the pressure-induced conformational fluctuations of the protein along with the free-energy landscape. The ground state of the protein has larger volume fluctuations around the cavities as compared to the rest of the protein, as is evident from the chemical shift changes. The ground state of L99A seems to be comprised of an ensemble of conformations in which the orientation of the methyl groups and the accommodation of water molecules in the cavity are both different. Elevated pressure stabilized the lower volume conformers in the ground state ensemble, resulting in the formation of a heterogeneous ensemble of differently solvated conformers. As the pressure-induced changes cause a chemical shift, as well as a chemical shift-dependent line-broadening, the volume fluctuations within the ground state ensemble would occur with rate constants similar to or faster than that of the NMR time scale ($\tau \ll$ ms).

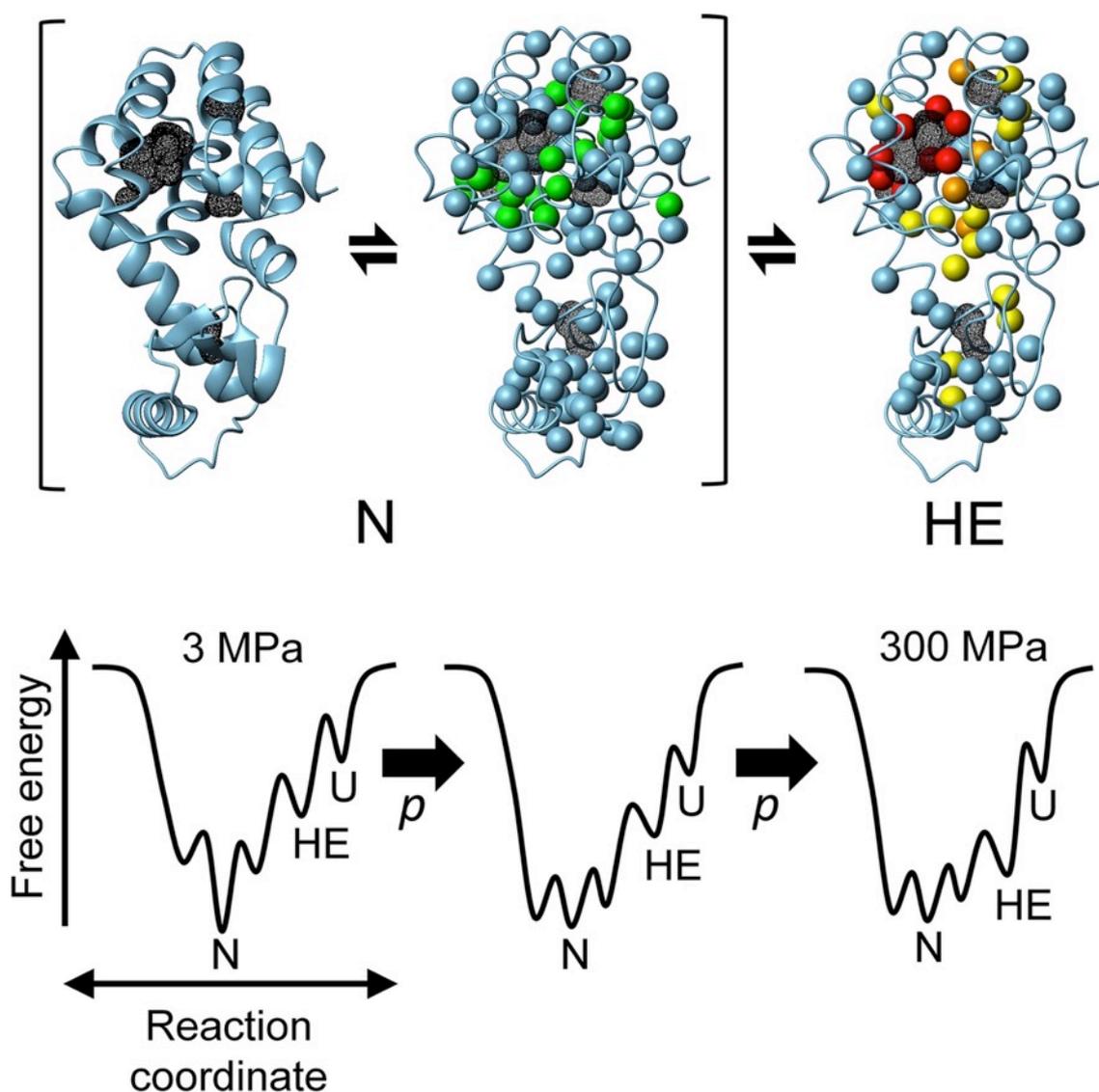

**FIGURE 7.** Schematic representation of the conformational equilibria in the T4 lysozyme L99A mutant along with the free-energy landscape. N, HE, and U represent the ground state ensemble, the high Gibbs free-energy state, and the unfolded state, respectively. The methyl groups expected to display a large volume fluctuation within the ground state ensemble are represented by green spheres (see Fig. 1*c*). Side-chain methyl groups classified into the rapid and intermediate decaying groups are represented by red and yellow spheres, respectively (see Fig. 2*b*).

In addition, the ground state ensemble is in equilibrium with a high-energy state (Fig. 7). We found that the structural characteristics of the pressure-stabilized state of L99A closely resemble those of the transiently formed high-energy state seen at atmospheric pressure, in which the F114 aromatic ring is accommodated in the enlarged hydrophobic cavity. This coincidence validates the argument that an elevated pressure stabilizes the intrinsically existing high-energy conformations of proteins at atmospheric pressure. Under high pressure conditions, the protein uses cavities to adopt a smaller partial molar volume, using either or both of the fluctuations.

A similar loss of the HSQC crosspeaks was also observed around the water-containing cavities in hen egg white lysozyme, when the temperature decreases at high pressure. The disappearance of NMR signals suggests the formation of a heterogeneous ensemble of partially solvated conformations, facilitated by the penetration of more water molecules into cavities of the protein (24). As the water-containing cavity is conserved among lysozymes across a variety of biological species, it seems that water-containing cavities play an important role in biological functions (e.g., allocating a certain degree of mobility to the active site and providing water molecules for hydrolysis reactions). After the atomic occupancy of the cavities becomes substantial owing to the hydration or ligand binding under high pressure, proteins usually collapse their cavities and unfold partially or completely, as seen in the case of ubiquitin (21), β-lactoglobulin (62), RalGDS-RBD (15), staphylococcal nuclease (25), and OspA (20). As structural changes are necessary for diverse functions such as ligand binding and signal transduction, the internal cavities of proteins could serve as an important structural element for a variety of biological events.

## ACKNOWLEDGEMENTS

We thank L. E. Kay and G. Bouvignies for providing us with the ZZ-exchange spectra of the L99A/G113A mutant of T4 lysozyme. This work was supported by a Grant-in-Aid for Scientific Research on Innovative Areas from the MEXT of Japan to R. K (23107729).

## SUPPORTING CITATIONS

Reference (64) appears in the Supporting Material.

# Supporting materials

**Methods**

**RISM theory**

3D-RISM is a computational solvent representation that is based on statistical liquid theory (1,2). We briefly review the equations below, as they have already been discussed in detail elsewhere (3,4). The 3D-RISM equation is given by:

$$h_\gamma^{uv}(\vec{r}) = \sum_{v'\in V} \sum_{\alpha \epsilon v'} c_\alpha^{uv'} * [\omega_{\alpha\gamma}^{v'v} + \rho^{v'} h_{\alpha\gamma}^{v'v}](\vec{r}) \qquad (1)$$

where here *h*, *c*, and *ω* are the total, direct and intramolecular correlation functions, respectively, and the asterisk denotes the convolution integrals. $\rho$ is the average density of solvent species, whereas *v*, *γ*, and *α* represent the solvent site of interest and reference site, respectively The term *V* represents all of the solvent species, whereas *u* and *v* represent the sites contained in the solute and solvent, respectively. In order to close the above equation, we choose the Kovalenko-Hirata closure (5):

$$g_\gamma^{uv}(\vec{r}) = \begin{cases} exp\left(d_\gamma^{uv}(\vec{r})\right) & for \quad d_\gamma^{uv}(\vec{r}) \le 0 \\ 1 + d_\gamma^{uv}(\vec{r}) & for \quad d_\gamma^{uv}(\vec{r}) > 0 \end{cases} \qquad (2)$$

$$g_\gamma^{uv}(\vec{r}) = h_\gamma^{uv}(\vec{r}) + 1$$

$$d_\gamma^{uv}(\vec{r}) = -\beta U_\gamma^{uv}(\vec{r}) + h_\gamma^{uv}(\vec{r}) - c_\gamma^{uv}(\vec{r})$$

where

$$U_\gamma^{uv}(\vec{r}) = \sum_{\alpha \in u} U_{\gamma\alpha}^{MM,NB}(\vec{r}-\vec{r}_{alpha})$$

where *γ* is the distribution function; *β*, the inverse temperature; and $U^{MM,NB}$, the atomic distance-dependent non-bonded molecular mechanics interaction potential used in typical simulations.

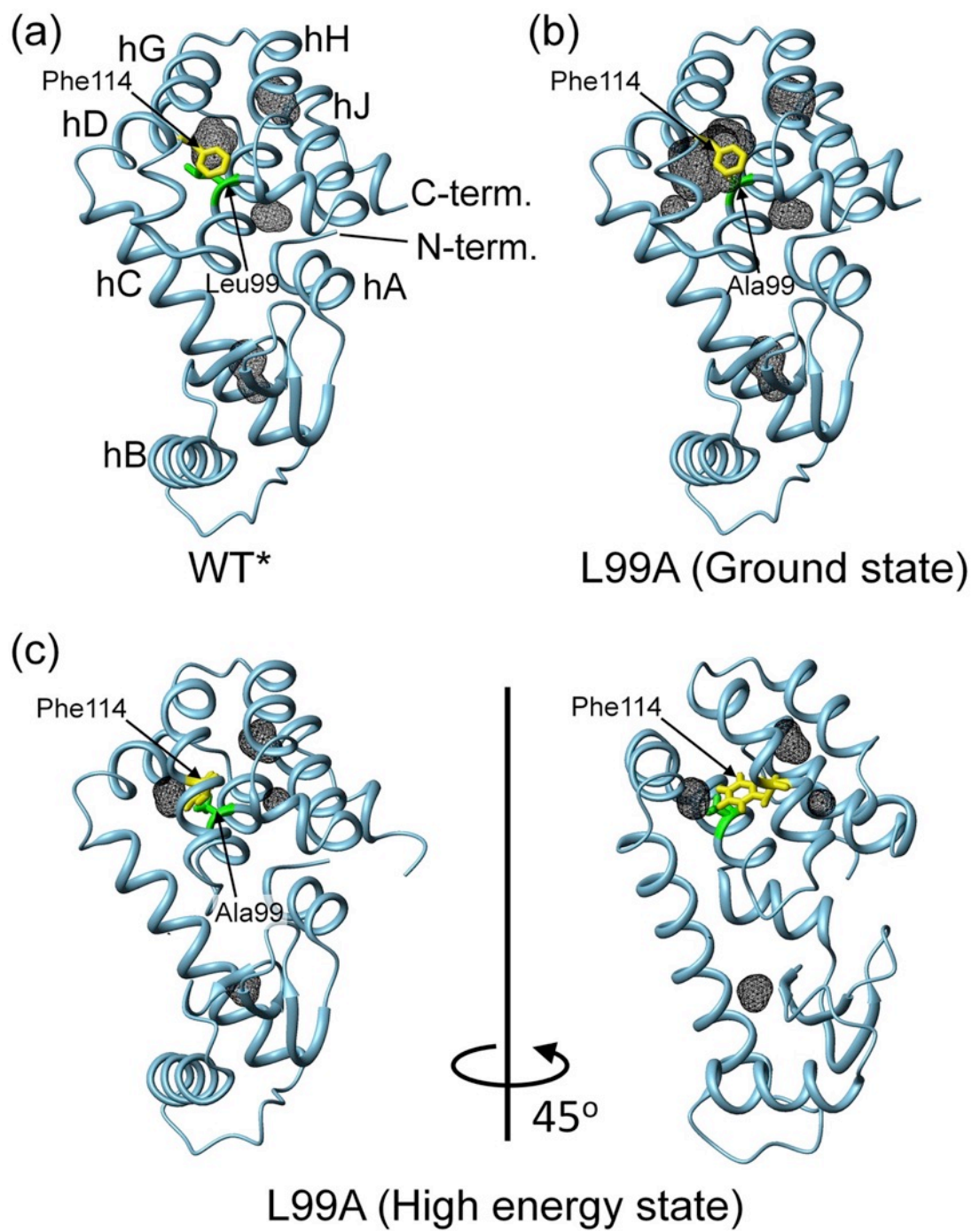

**Fig. S1.** Three-dimensional structure of T4 lysozyme, (*a*) the pseudo wild-type WT* (PDB ID: 6LZM) and (*b*) L99A (PDB ID: 1L90), and (*c*) L99A high-energy state (PDB ID: 2LCB). Internal cavities are represented by black wire-frame cages calculated by a program MOLMOL with a probe radius of 1.4 Å. Residues 99 and 114 are indicated by stick representations.

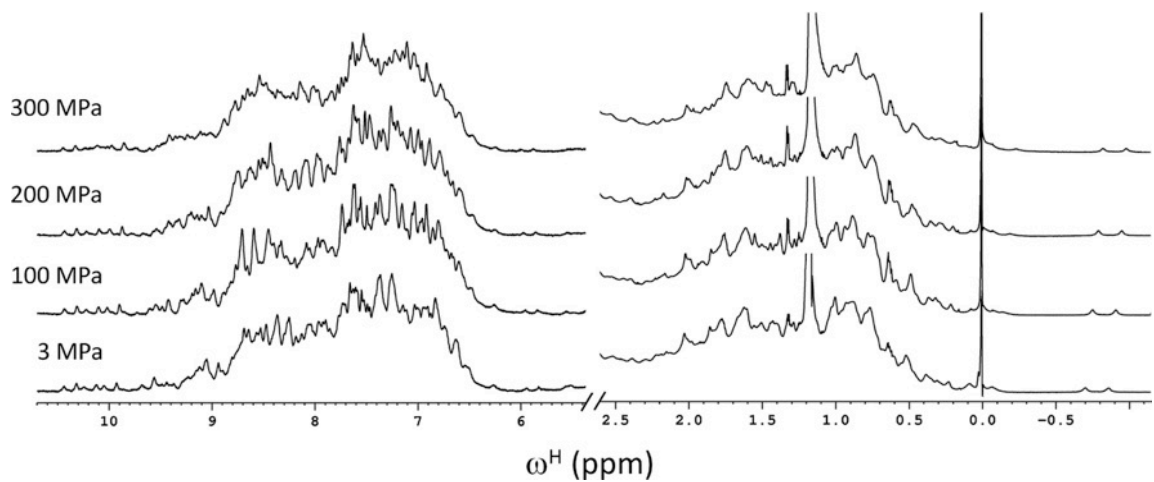

**Fig. S2.** One-dimensional $^1$H NMR spectra of uniformly $^{15}$N/$^{13}$C-labeled L99A protein at different pressures from 3 to 300 MPa at 25 °C. NMR peaks from protons attached to $^{13}$C or $^{15}$N are split due to J-coupling.

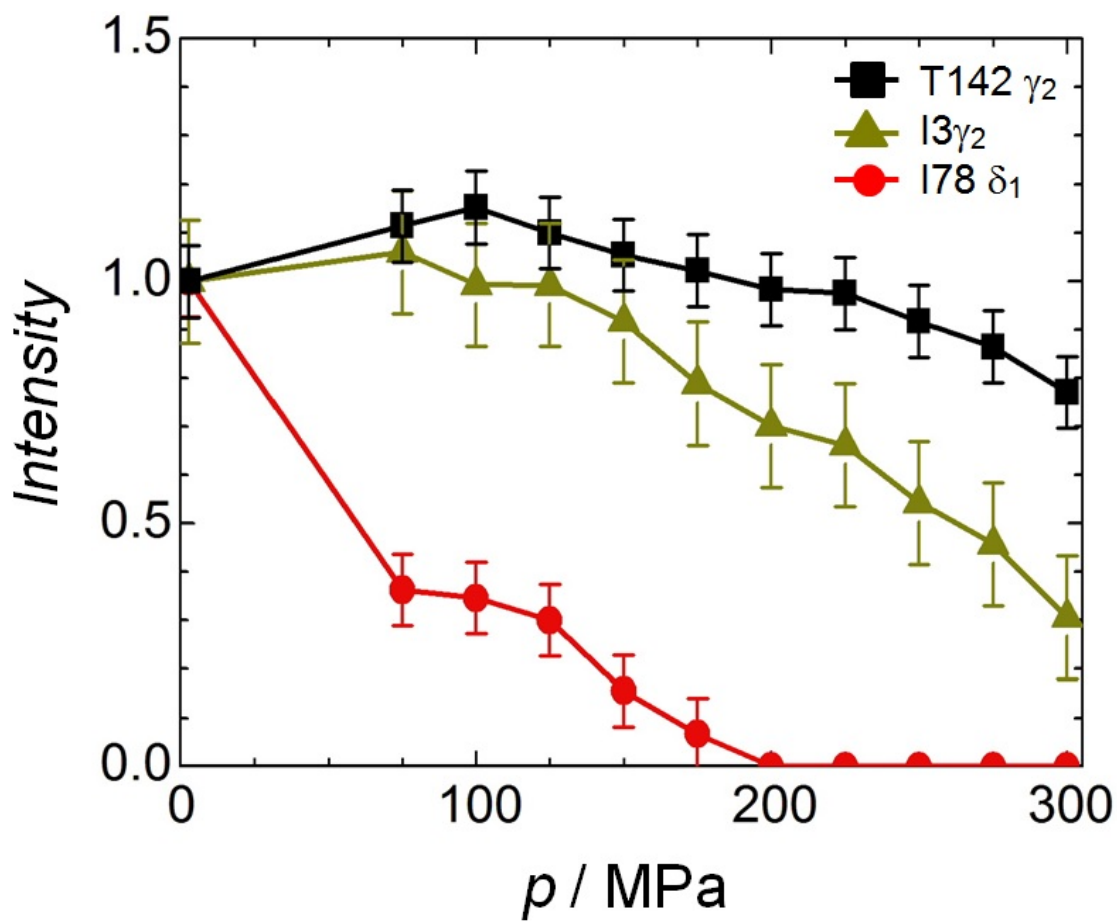

**Fig. S3.** Errors in peak intensity (i.e. maximum peak height) estimation for representative methyl groups, T142$\gamma_2$, I3$\gamma_2$, and I78$\delta_1$, which belong to the slow, intermediate, and rapidly decaying groups, respectively (see Fig. 2). Error bars shows the standard deviation of the noise, relative to the signal intensity at 3 MPa.

(a)

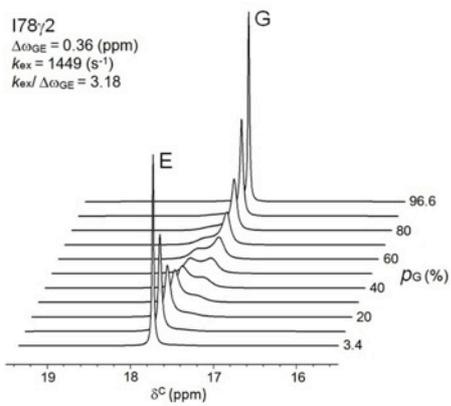
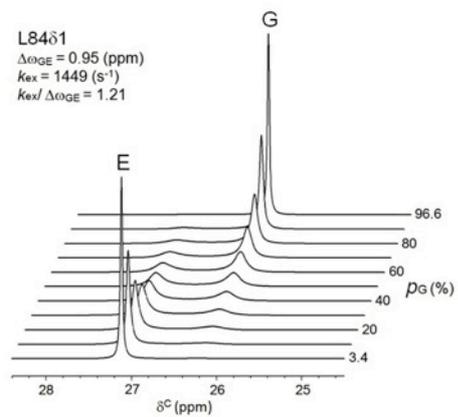
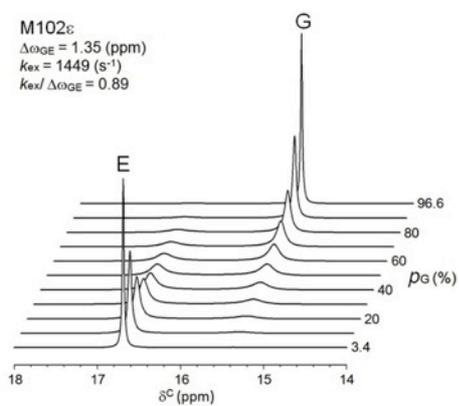
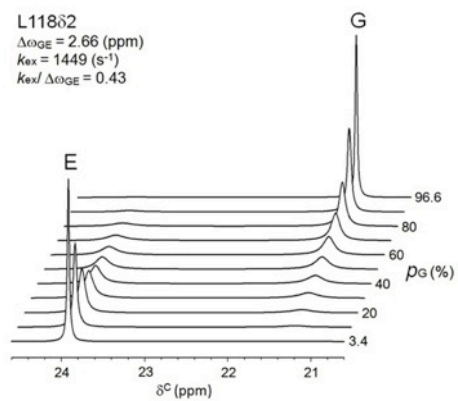
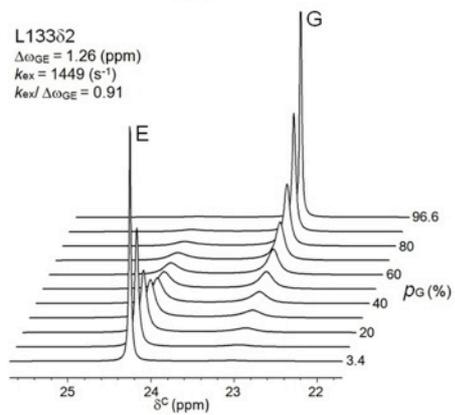

(b)

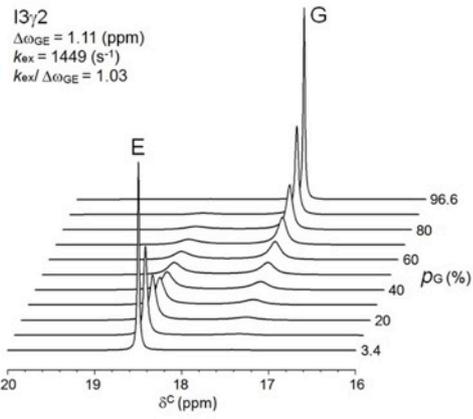
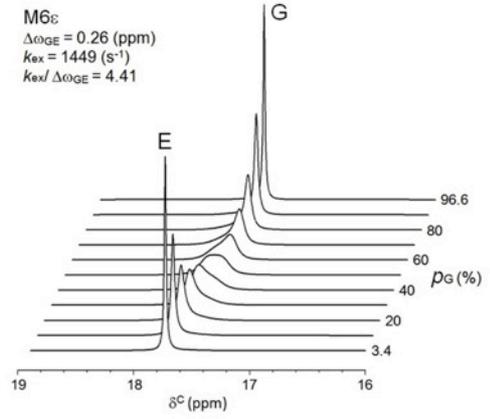
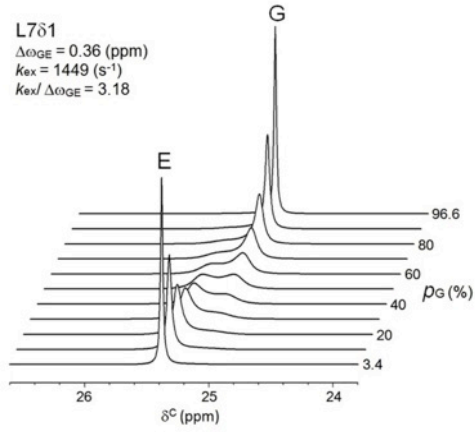
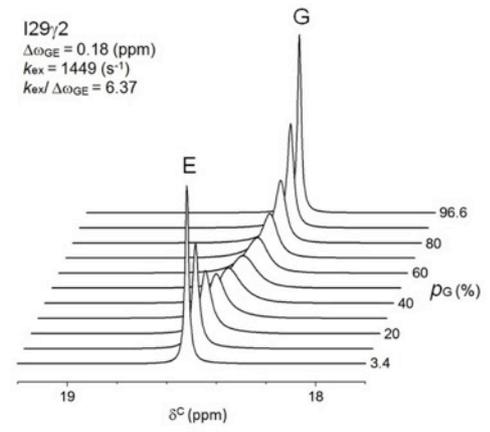
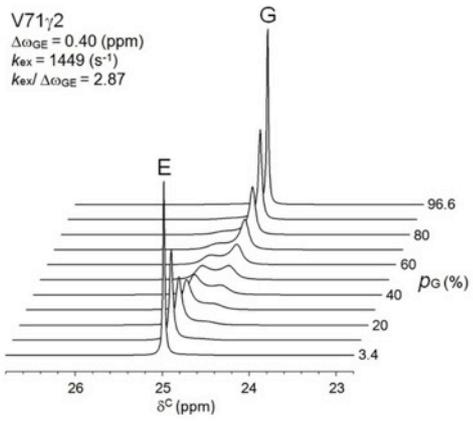
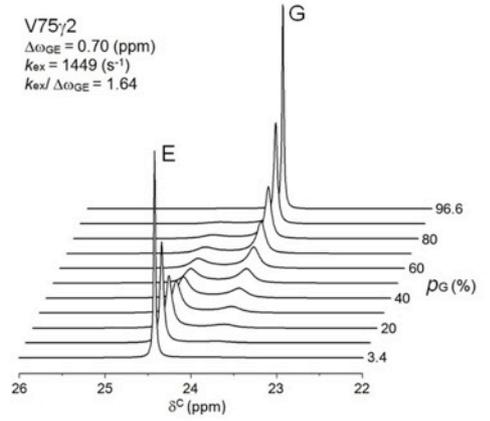

(b)

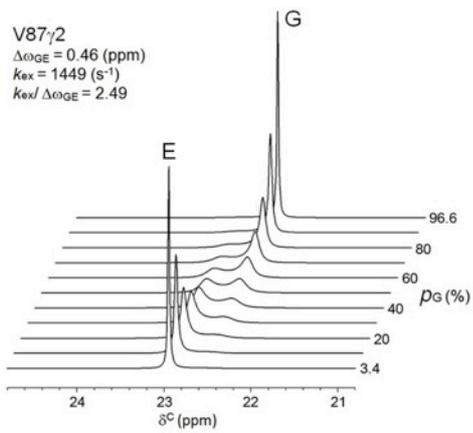
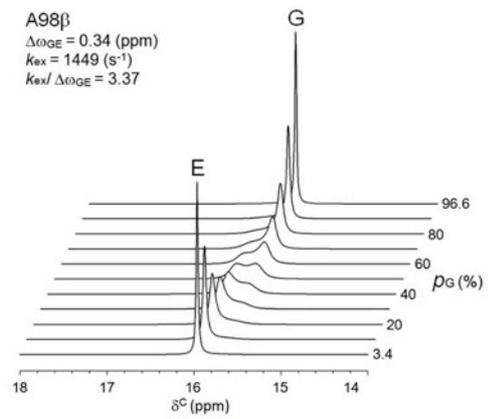
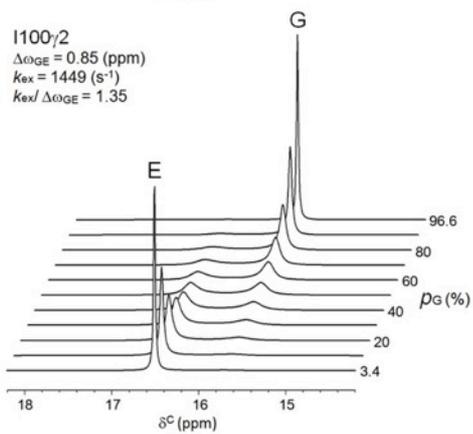
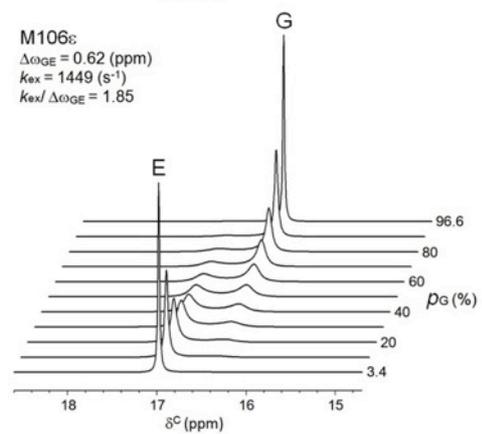
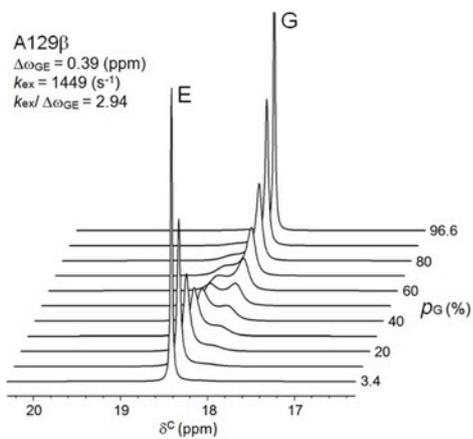
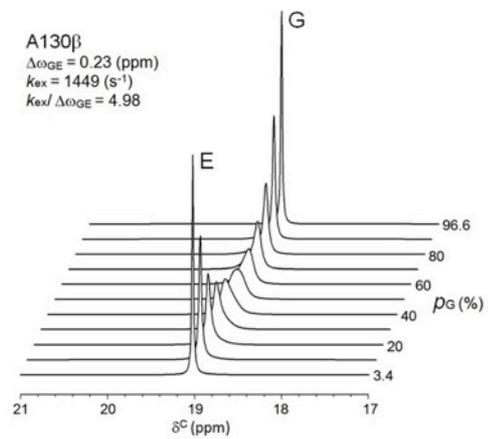

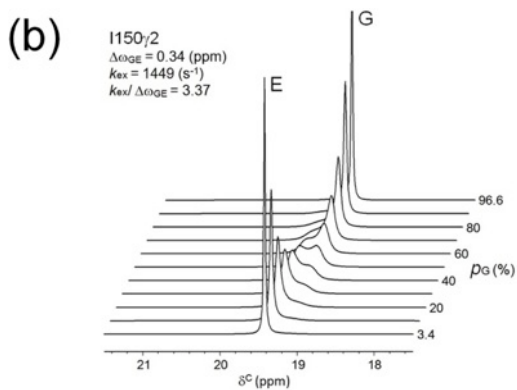

**Fig. S4.** Line-shape simulation for the methyl carbons of L99A involved in (*a*) the rapidly decaying group (e.g. I78γ$_2$, L84δ$_1$, M102ε, L118δ$_2$ and L133δ$_2$) and (*b*) the intermediate decaying group (e.g. I3γ$_2$, M6ε, L7δ$_1$, I29γ$_2$, V71γ$_2$, V75γ$_2$, V87γ$_2$, A98β, I100γ$_2$, M106ε, A129β, A130β, and I150γ$_2$) at different population of the high-energy conformer with an assumption of the two-state exchange model. Chemical shift difference Δω, time constant $k_{ex}$ of chemical exchange between the ground state and the transiently populated high-energy state, and a population of the high-energy state $p_E$ were all obtained from the literature (6). Chemical shifts of the ground state of L99A were collected from $^1$H/$^{13}$C-HSQC spectrum obtained at 3 MPa and 25 °C (Fig. 1*a*). Several residues (I78δ$_1$ and A99β which belong to the rapid decaying group and I3δ$_1$, I29δ$_1$, L46δ$_1$, L66δ$_1$, I150δ$_1$ which belong to the intermediate decaying group) are excluded from this analysis because the parameters for line-shape simulations (Δω, $k_{ex}$ and $p_E$) were not obtained in the previous NMR $^{13}$C-$R_2$ dispersion experiment (6). The program WINDNMR-Pro was used for the simulations.

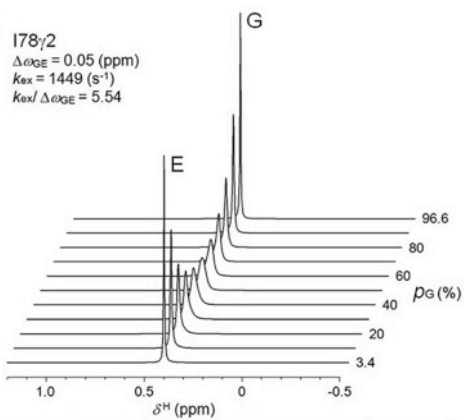
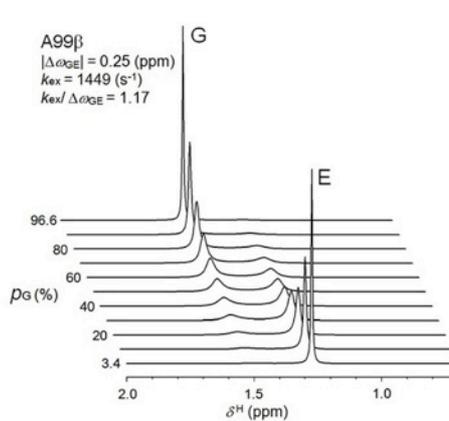
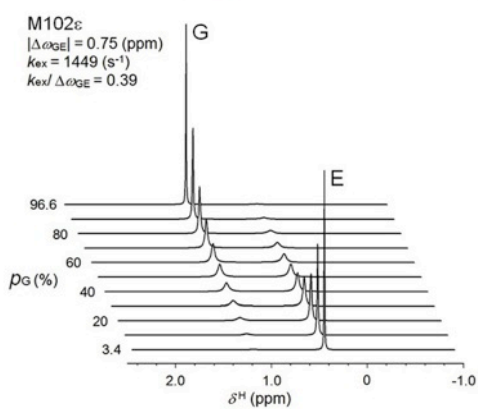
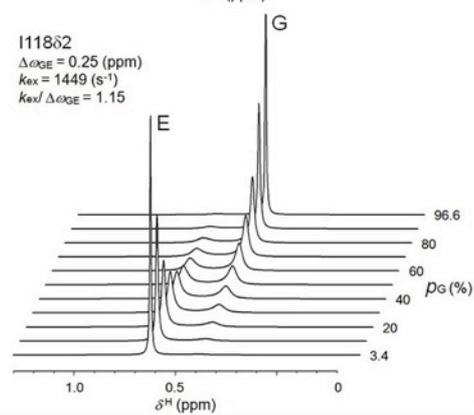

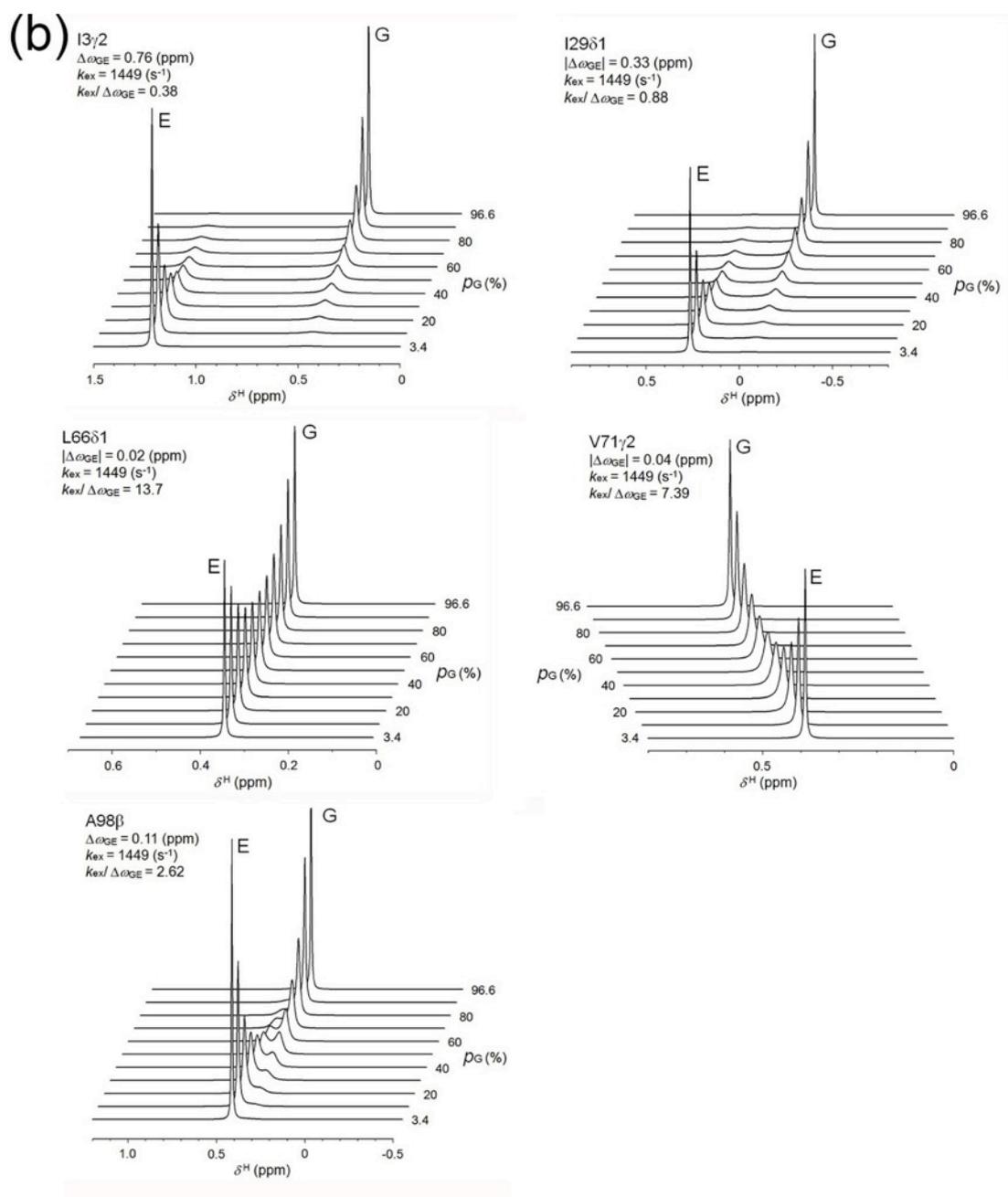

**Fig. S5.** Line-shape simulation for the methyl protons of L99A involved in (*a*) the rapidly decaying group (e.g. I78Hγ$_2$, A99Hβ, M102Hε, and I118Hδ$_2$) and (*b*) the intermediate decaying group (e.g. I3Hγ$_2$, I29Hδ$_1$, L66Hδ$_1$, V71Hγ$_2$, and A98Hβ) at different population of the high-energy conformer with an assumption of the two-state exchange model. Chemical shift differences Δω between the ground state and the high-energy state of the L99A/G113A mutant were used in the simulation.

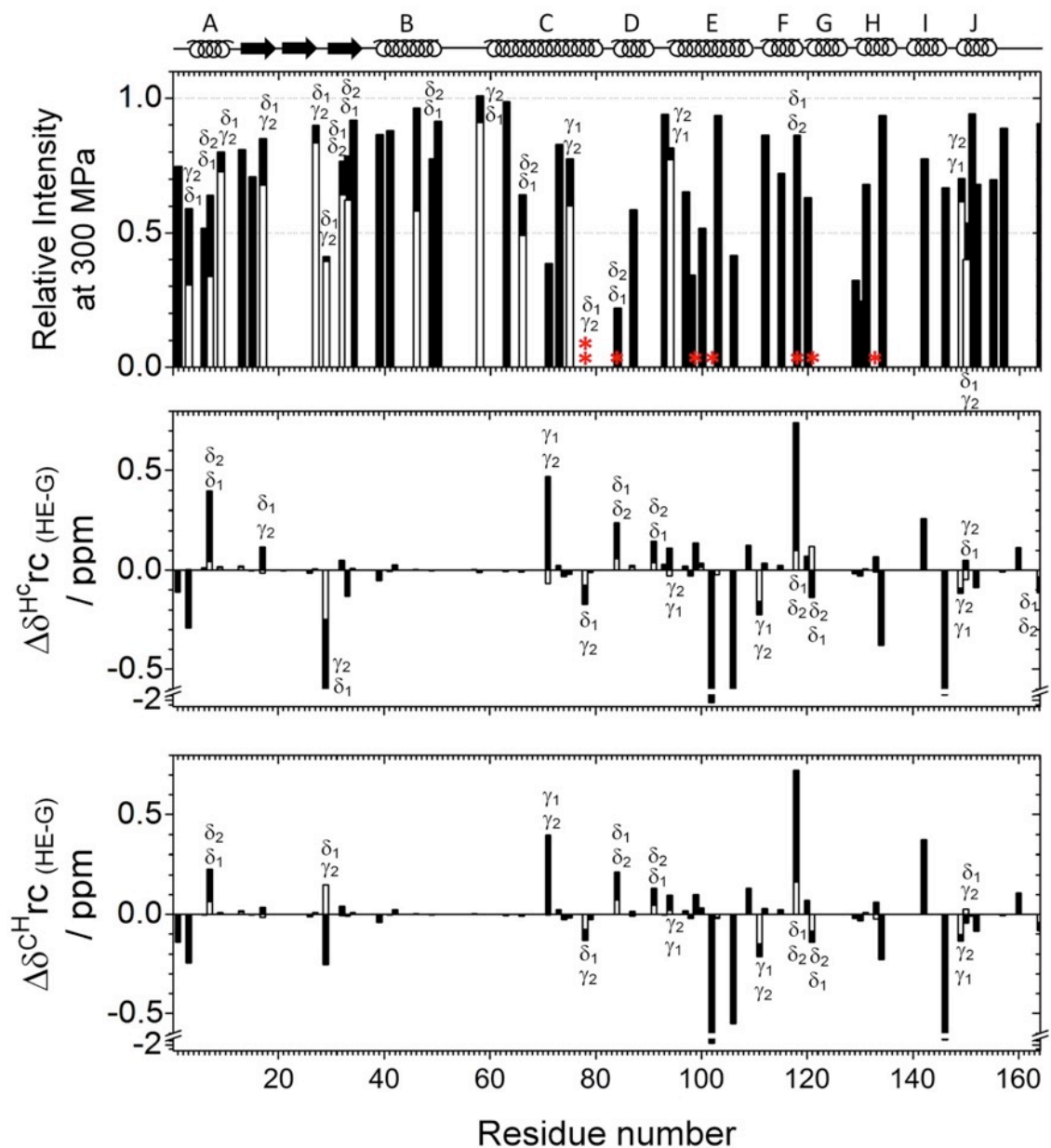

**Fig. S6.** (*a*) The relative intensities of $^1$H/$^{13}$C HSQC cross-peaks for L99A at 300 MPa, normalized by those at 0.1 MPa, alongside residue number. Methyl groups showing zero intensity at 300 MPa are marked by asterisks. Secondary structures are represented at the top of each panel (rings for α-helix, arrows for β-stand). Dotted lines show a relative intensity of 0.5. (*b*) Differences of $^1$H (*top*) and $^{13}$C (*bottom*) ring current shifts between the ground (G) and high-energy (HE) states ($\Delta\delta_{rc}$) for the methyl groups along with the residue number. Residues showing remarkable changes are shown by residue numbers. When the residue has two methyl groups, the group indicated at the bottom shows a larger absolute value than the top.

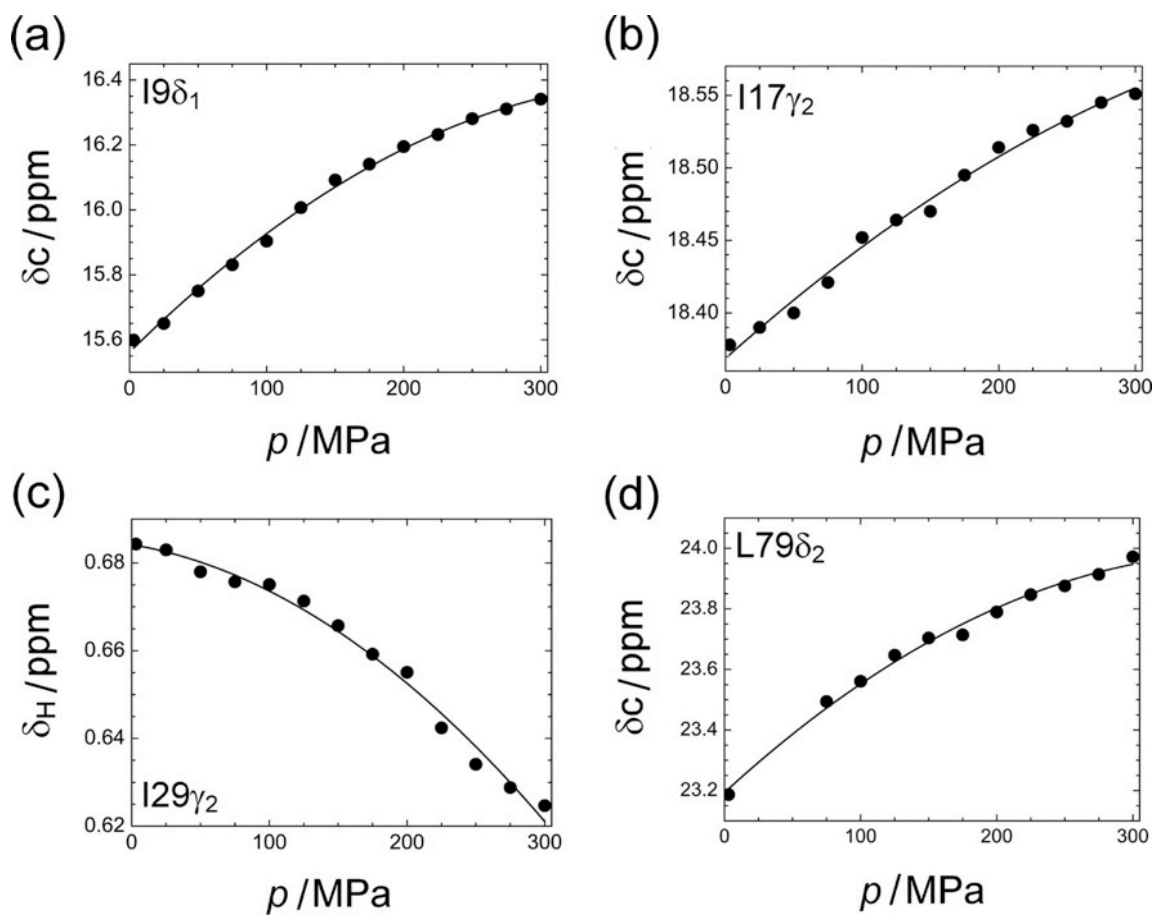

**Fig. S7.** Non-linear pressure-induced chemical shifts changes for I9 C$\delta_1$ (*a*), I17 C$\gamma_2$ (*b*), I29 H$\gamma_2$ (*c*), and L79 C$\delta_2$ (*d*). Some data for L79$\delta_2$ are missing because of peak overlap.

**Supporting References**